\documentclass[fleqn,usenatbib]{mnras}
\usepackage{longtable}
\usepackage{newtxtext,newtxmath}

\usepackage[T1]{fontenc}

\DeclareRobustCommand{\VAN}[3]{#2}
\let\VANthebibliography\thebibliography
\def\thebibliography{\DeclareRobustCommand{\VAN}[3]{##3}\VANthebibliography}

\usepackage{graphicx}	
\usepackage{amsmath}	
\usepackage{pdflscape}

\usepackage[dvipsnames]{xcolor}            
\usepackage{ulem}                          
\title[Gaia EDR3 Pleiades]{Reconstructing the Pleiades with Gaia EDR3}

\author[Heyl, Caiazzo \& Richer]{
Jeremy Heyl\thanks{email: heyl@phas.ubca.ca}$^1$,
Ilaria Caiazzo\thanks{email: ilariac@caltech.edu; Sherman Fairchild Fellow}$^2$, Harvey Richer$^1$
\\
$^{1}$Department of Physics and Astronomy, University of British Columbia, Vancouver, BC V6T 1Z1, Canada\\
$^{2}$TAPIR, Walter Burke Institute for Theoretical Physics, Mail Code 350-17, Caltech, Pasadena, CA 91125, USA\\
}

\date{Accepted XXX. Received YYY; in original form ZZZ}

\pubyear{2021}

\begin{document}
\label{firstpage}
\pagerange{\pageref{firstpage}--\pageref{lastpage}}
\maketitle

\begin{abstract}
We search through an eight-million cubic-parsec volume surrounding the Pleiades star cluster and the Sun to identify both the current and past members of the Pleiades cluster within the Gaia EDR3 dataset.  We find nearly 1,300 current cluster members and 289 former cluster candidates. Many of these candidates lie well in front or behind the cluster from our point of view, so formerly they were considered cluster members, but their parallaxes put them more than 10~pc from the centre of the cluster today. Over the past 100~Myr we estimate that the cluster has lost twenty percent of its mass including two massive white dwarf stars and the $\alpha^2$ Canum Venaticorum-type variable star, 41 Tau.  All three white dwarfs associated with the cluster are massive ($1.01-1.06~\textrm{M}_\odot$) and have progenitors with main-sequence masses of about six solar masses.  Although we did not associate any giant stars with the cluster, the cooling time of the oldest white dwarf of 60~Myr gives a firm lower limit on the age of the cluster.
\end{abstract}



\section{Introduction}

The Pleiades have fascinated humans for millennia.  As the most prominent open cluster close to Earth, the Pleiades has long provided a benchmark for our understanding of stellar evolution and dynamics.  The cluster also contributed the first rung on the cosmological distance ladder by providing a single-age stellar population with measured parallaxes \citep{1994MNRAS.266..441L,1997AJ....113...13B,1999PASP..111..775F}. The first parallax measurements with Hipparcos, though, resulted in a tension for stellar modelling with an unexpected large parallax for the cluster \citep{1997ESASP.402..643M,2001A&A...374..105S}. Gaia DR2 \citep{2018A&A...616A...1G} and Gaia EDR3 \citep{2020arXiv201201533G}, with their more precise parallax measurements, recently resolved this tension \citep{2018A&A...616A..10G}, and the cluster stars now show excellent agreement with the expected magnitudes of a stellar population with a metallicity close to solar and an age of about 100~Myr \citep{2018A&A...616A..10G}. Unfortunately, up to now, only a single evolved star, the white dwarf EGGR 25, has been associated with the Pleiades \citep{1965ApJ...141...83E}.  The properties of white dwarfs and giant stars can yield more precise age estimates from the white-dwarf cooling \citep{2013Natur.500...51H} and turn-off techniques \citep{1999AJ....118.2306R,2005AJ....130..116D,2011ApJ...738...74D}. 

Several recent works \citep{2019A&A...623A..35L,2019A&A...628A..66L,2021A&A...645A..84M} have examined nearby stellar clusters using data from Gaia DR2 following in spirit of the pioneering work of \citet{1998A&A...331...81P} using data from Hipparcos. Although we do identify the Pleiades cluster in this work and find similar properties to those found by \citet{2019A&A...628A..66L,2021A&A...645A..84M}, the identification of the cluster itself and its centre of mass is just the first step in our analysis which is to also find the former members of the Pleiades.  We perform our search in the fully six-dimensional phase space by reconstructing the required radial velocity to bring each observed star as close to the Pleiades as possible in its past. To be a candidate escapee from the Pleiades the distance of closest approach must be less than fifteen parsecs to allow for the radius of the cluster (about 10~pc) and include for some error in the distance determination of a few parsec, the relative velocity must be less than 2.4~km/s (this is typically of escapees) and the time of closest approach must be within the lifetime of the cluster.  For many stars, we can compare the reconstructed radial velocity with the one observed with Gaia, providing an additional check of our technique.  For the recently departed, the stars are still near the cluster and form a corona \citep[\textit{c.f.}][]{2021A&A...645A..84M}, but the more distantly departed are spread throughout the sampling volume and presumably beyond as well.


With these issues in mind and the release of more accurate proper motion and parallax measurements of Gaia EDR3 \citep{2020arXiv201201533G}, we have searched the Gaia EDR3 catalogue for potential escapees from the Pleiades cluster.  We outline our geometric technique for finding candidate escapees in \S~\ref{sec:methods}, the properties of the escapee population in \S~\ref{sec:results} and the implications of the white-dwarf escapees in \S~\ref{sec:white-dwarfs}.  We present the details of constructing the dataset and the entire catalogue of cluster members and escapee candidates in appendices.

\section{Methods}
\label{sec:methods}

\begin{figure*}
    \centering
    \includegraphics[width=0.32\textwidth,trim=0 0.2in 0 0]{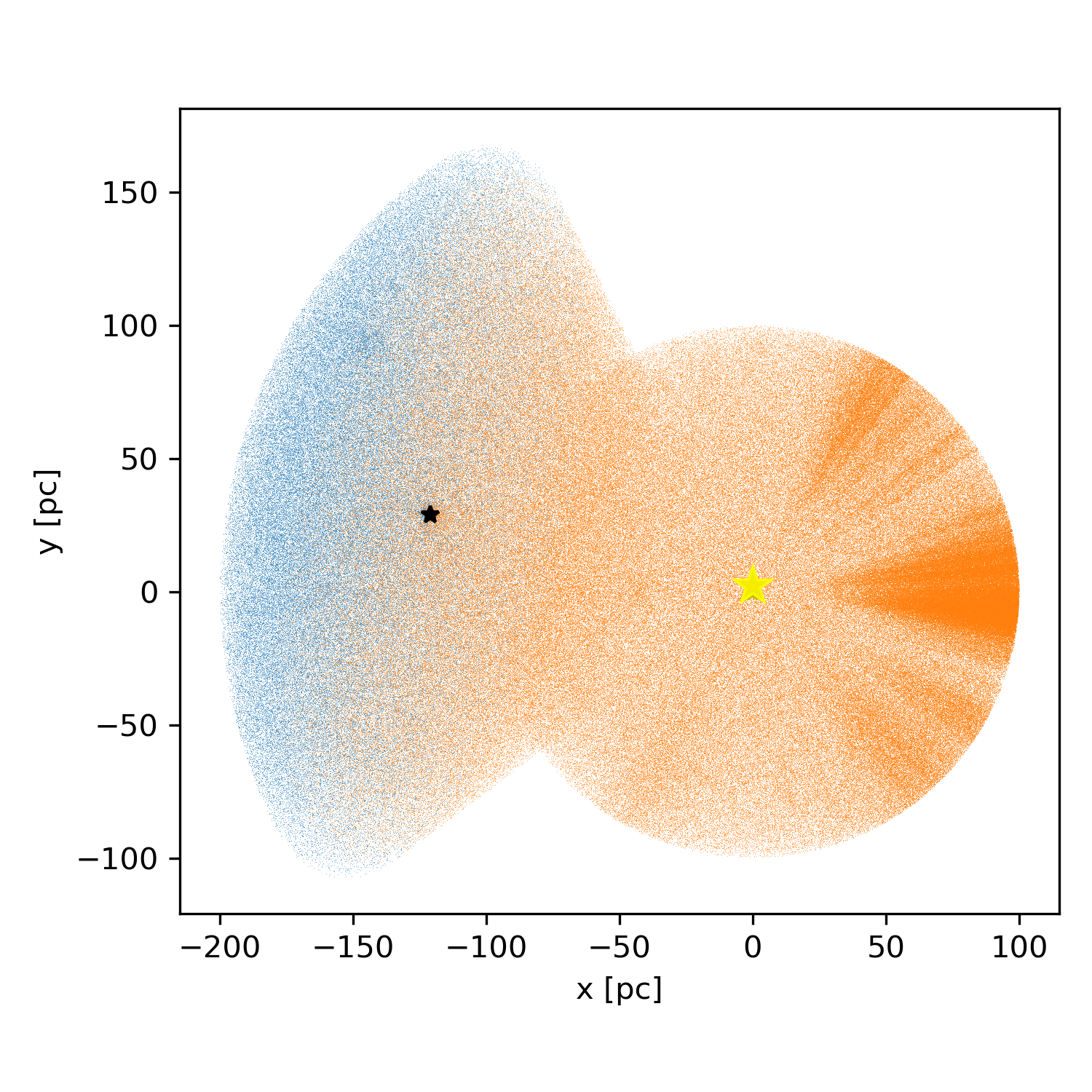}
    \includegraphics[width=0.32\textwidth,trim=0 0.2in 0 0]{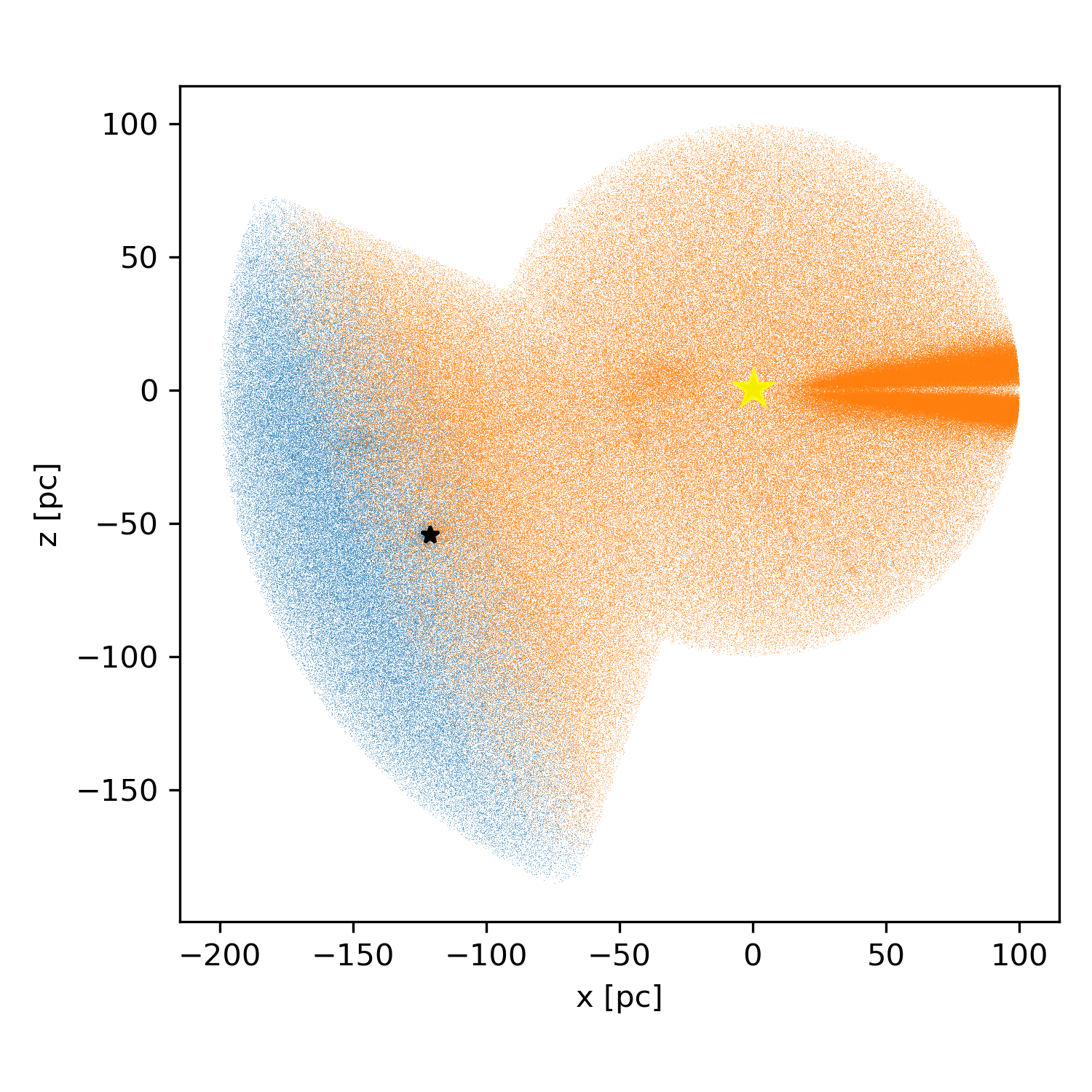}
    \includegraphics[width=0.32\textwidth]{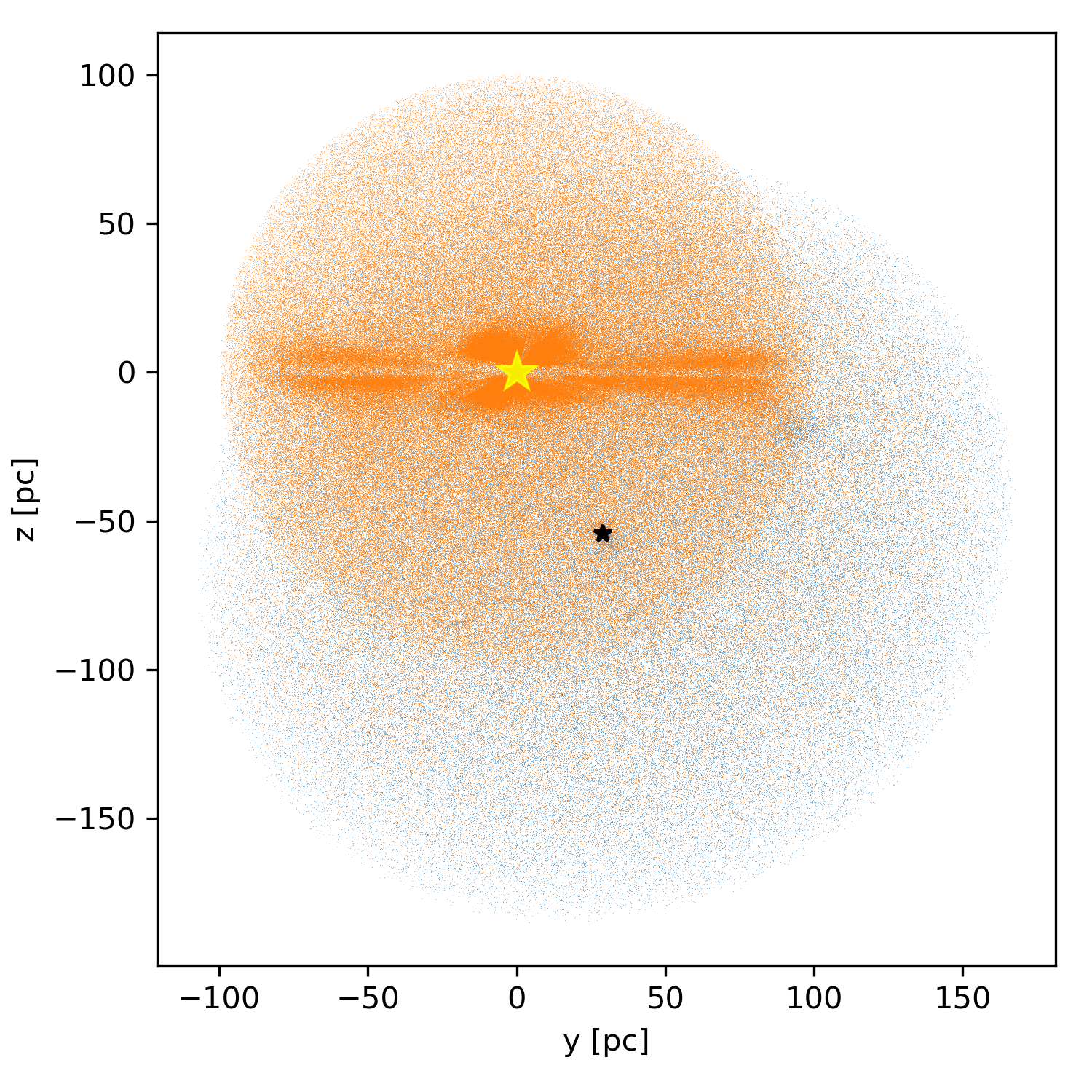}
    
    \caption{The Gaia EDR3 sample.  The blue objects lie beyond the plane containing the Pleiades (indicated with a black star) and perpendicular to the line of sight from the Sun which is indicated with a yellow star at (0,0,0).}
    \label{fig:positions}
\end{figure*}

We construct our Gaia EDR3 sample by finding all objects within 100~pc of the Sun and additionally all objects within 200~pc of the Sun that lie within 45~degrees of the Pleiades on the sky.  This results in a four-million-cubic parsec sphere centred on the Sun and another contiguous four-million-cubic parsec region centred on the Pleiades as shown in Fig.~\ref{fig:positions}.  To analyse the sampled volume in greater detail, we divide it into a region beyond the Pleiades (blue) and one closer to the Sun (orange).  From the point of view of the Pleiades itself, which is about 136~pc away from the Sun, the sample contains the entire volume within 64~pc of the Pleiades, as well as an entire hemisphere of radius 96~pc on the nearside.

\begin{figure}
    \centering
    \includegraphics[width=\columnwidth]{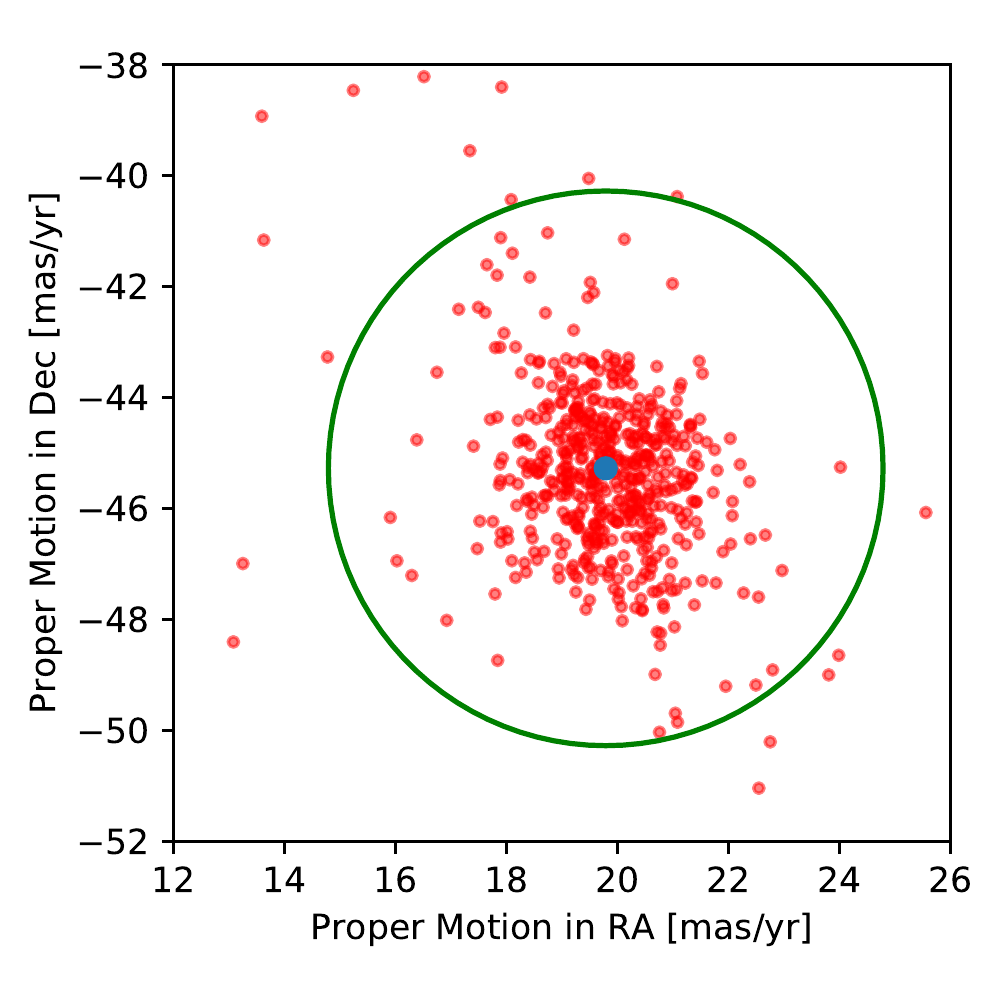}
    \caption{Proper motion of stars within one degree of the centre of the Pleiades.  The median proper motion and the five milliarcsecond per year selection region are also depicted.}
    \label{fig:pmselection}
\end{figure}

\begin{figure}
    \centering
    \includegraphics[width=\columnwidth]{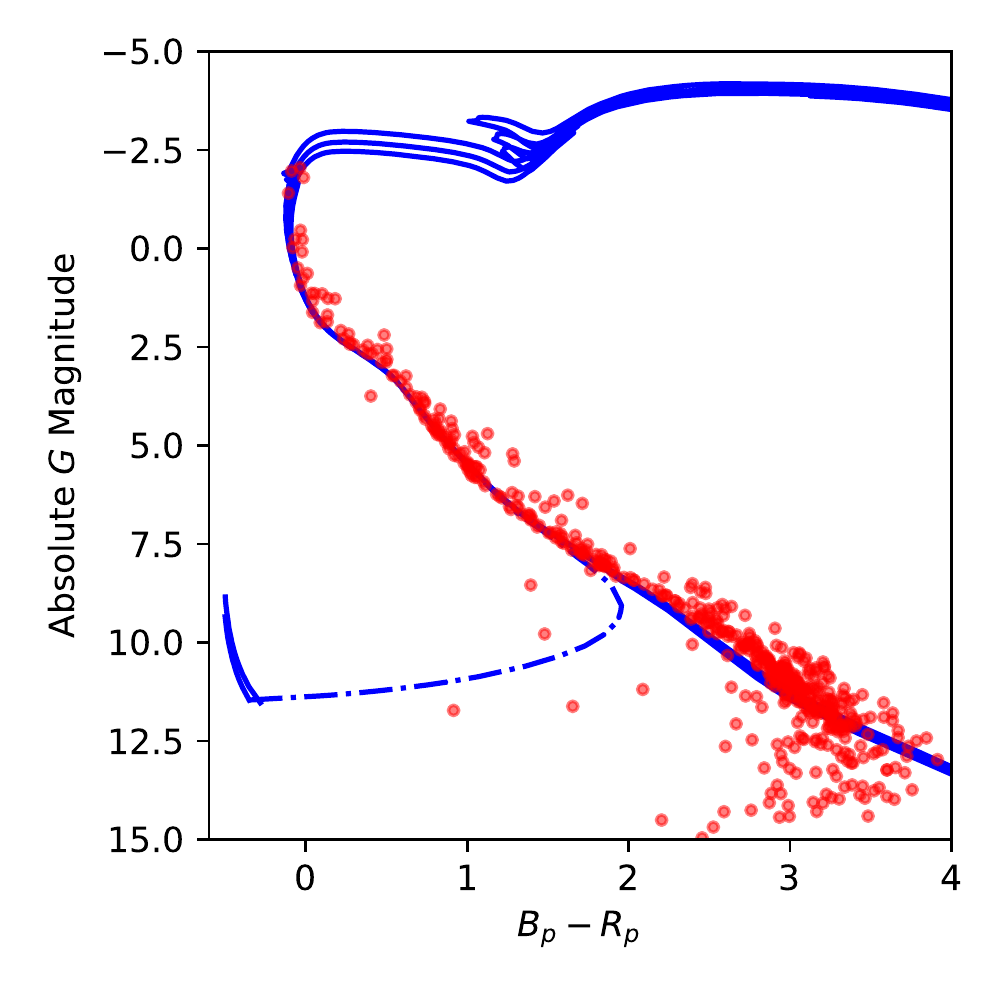}
    \caption{Gaia EDR3 colour-magnitude diagram of the proper-motion selected sample.  The blue curves are Padova isochrones with ages of 110, 130 and 150 Myr and Montreal white-dwarf models of 1.0 and 1.1 solar masses up to a cooling age of 110~Myr.  An extinction of $E(B_p-R_p)=0.08$ and $A_G=0.17$ has been applied to the models.  To convert the apparent $G$ magnitudes to absolute, we use the median distance of the sample of 136~pc obtained from the parallaxes. A binary composed the oldest white dwarf expected in the cluster and a main-sequence star is shown by the dot-dashed curve. }
    \label{fig:cmd=pmselection}
\end{figure}

To find the centre of mass motion and position for the Pleiades, we examine the proper motions of all stars within one degree on the sky from the centre of the cluster as shown in Fig.~\ref{fig:pmselection}.   We iteratively calculate the median proper motion of the cluster by starting with an initial guess of $(19, -45)~\textrm{mas/yr}$ and considering only the proper motions of stars within five milliarcseconds per year of the median, yielding the median proper motion and sample depicted in the figure.  The colour-magnitude diagram (CMD) of this sample is depicted in Fig.~\ref{fig:cmd=pmselection}.  We see that the cluster is well characterized by Padova isochrones from 100 to 140~Myr in age \citep{2012MNRAS.427..127B,2014MNRAS.445.4287T,2014MNRAS.444.2525C,2015MNRAS.452.1068C,2017ApJ...835...77M,2019MNRAS.485.5666P,2020MNRAS.498.3283P}.  On the same colour-magnitude diagram we have also plotted Montreal white-dwarf cooling models  up to a cooling age of 110~Myr \citep{1995PASP..107.1047B,2006ApJ...651L.137K,2006AJ....132.1221H,2011ApJ...730..128T,2011ApJ...737...28B,2018ApJ...863..184B,2020ApJ...901...93B}.  The white dwarf EGGR 25 does not lie within this sample as it is located more than a degree away from the cluster's centre.

Using those stars within the sample that have radial velocities measured with Gaia DR2, we calculate the mean velocity of the stars within the Pleiades relative to the Sun as
\begin{equation}
 {\bf v}_\textrm{cluster} = (-7.1\pm 0.4, -28.3 \pm 0.1,-14.4 \pm 0.2 )~\textrm{km/s}
 \label{eq:1}
 \end{equation}
 in Galactic coordinates.  Using all of the stars within the sample, we calculate the mean displacement of the cluster relative to the Sun as 
 \begin{equation}
 {\bf r}_\textrm{cluster} = (-120.99\pm 0.20, 28.96 \pm 0.06,   -54.23 \pm 0.10)~\textrm{pc},
 \label{eq:2}
 \end{equation}
 again in Galactic coordinates.   Both of these values agree within the uncertainties with those derived by \citet{2019A&A...628A..66L} from Gaia DR2 data.
 
 Having determined the centre of the cluster in phase space, we will now used a more relaxed definition of the cluster sample to be all of the stars within 10~pc of the cluster centre whose proper motions lie within five milliarcseconds per year of the cluster median.  This 10~pc distance is similar to the tidal radii found by  \citet{2019A&A...628A..66L} and \citet{2021A&A...645A..84M} of 11.6 and 11.8~pc. This extended sample spans about eight degrees on the sky and its CMD is depicted in Fig.~\ref{fig:cmd}.  Furthermore, we find a similar number of cluster members within this volume of 1,272 stars, versus the 1,248 and 1,177 found by \citet{2019A&A...628A..66L} and \citet{2021A&A...645A..84M} respectively, and a similar total stellar mass of 627 solar masses compared to the 617.8 solar masses from \citet{2021A&A...645A..84M}.

To look for potential escapees from the Pleiades, we calculate the relative velocity within the plane of the sky, with respect to the cluster, of all of the stars in the sample as 
\begin{equation}
\Delta {\bf v}_\textrm{2D} = {\bf v}_\textrm{2D} - {\bf v}_\textrm{cluster} + \frac{ {\bf v}_\textrm{cluster} \cdot {\bf r}}{{\bf r} \cdot {\bf r}}{\bf r},
\label{eq:3}
\end{equation}
where ${\bf v}_\textrm{2D}$ is the velocity of the star in the plane of the sky, or its total velocity assuming zero radial velocity (see Appendix~A for the calculation).  
If we assume that stars escape from the cluster with a velocity of only a few kilometres per second and have not been accelerated subsequently, a small value of $\Delta {\bf v}_\textrm{2D}$ is a necessary but not sufficient condition to be a candidate escapee.  In fact, many objects have been identified as common-proper-motion partners of the Pleiades such as AB Dor \citep{2004ApJ...613L..65Z,2007MNRAS.377..441O} and the massive white dwarf GD~50 \citep{2006MNRAS.373L..45D,2018ApJ...861L..13G} (we find that the value of $\Delta {\bf v}_\textrm{2D}$ for GD~50 is only 200~m/s), but these objects were unlikely to be part of the cluster in the past as they are moving toward the cluster today.

\begin{figure}
    \centering
    \includegraphics[width=\columnwidth]{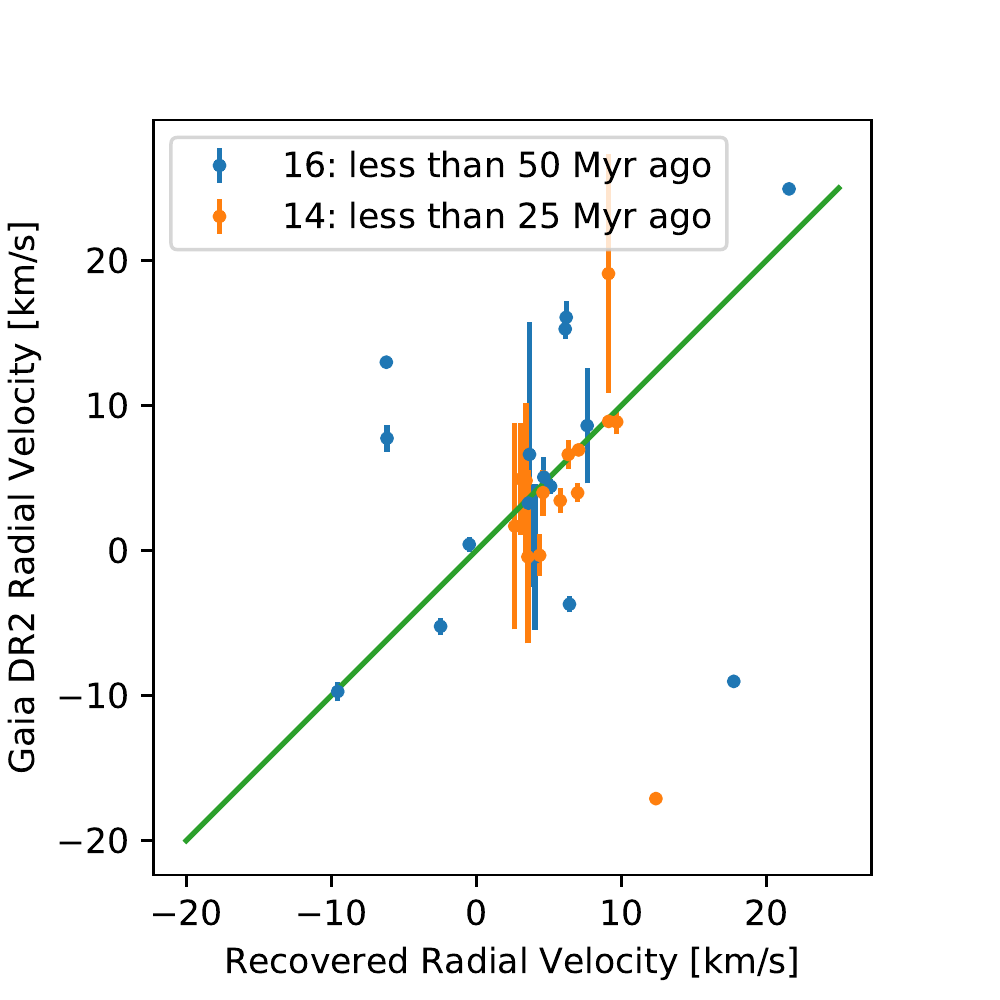}
    \caption{A comparison of Gaia DR2 radial velocities to the reconstructed radial velocities for escapee candidates.  }
    \label{fig:rvcomp}
\end{figure}

In fact, having a small proper motion relative to the cluster is not sufficient to be identified as a candidate escapee: an object must also have a relative proper motion away from the Pleiades that is sufficiently large that it could have reached its current location relative to the cluster within the lifetime of the cluster.  To find this sample, we determine the distance of each star from the cluster as a function of time assuming no acceleration and an arbitrary radial displacement ($\delta r$) as
\begin{equation}
d(t)^2 = \left [ {\bf r}-{\bf r}_\textrm{cluster} 
+  t \left ({\bf v}_\textrm{2D}-{\bf v}_\textrm{cluster} \right ) + \hat{\bf{r}} \delta r  \right ]^2,
\label{eq:4}
\end{equation}
and we look for the time when the star and the cluster were or will be closest together,
\begin{equation}
t_\textrm{min} = \frac{ \Delta {\bf r}\cdot  \Delta {\bf v} - \left (  \Delta {\bf r} \cdot \hat{\bf{r}}  \right ) \left (  \Delta {\bf v} \cdot \hat{\bf{r}}  \right ) }{\left (  \Delta {\bf v} \cdot \hat{\bf{r}}  \right )^2-\left (  \Delta {\bf v}\right )^2}
\label{eq:5}
\end{equation}
where
\begin{equation}
 \Delta {\bf r} = {\bf r}-{\bf r}_\textrm{cluster} ~\textrm{and}~
 \Delta {\bf v} = {\bf v}_\textrm{2D}-{\bf v}_\textrm{cluster},
 \label{eq:6}
 \end{equation}
This also yields an estimate of the radial displacement and velocity of the star
\begin{equation}
\delta r = v_r t_\textrm{min} =  -\hat{\bf{r}} \cdot \left ( \Delta {\bf r} +t_\textrm{min} \Delta {\bf v}   \right ) 
\label{eq:7}
\end{equation}
 so
 \begin{equation}
 \hat {\bf v}_\textrm{3D} = {\bf v}_\textrm{2D} + v_r \hat{\bf{r}}
 \label{eq:8}
 \end{equation}
 and
 \begin{equation}
 \Delta \hat {\bf v}_\textrm{3D} = \hat {\bf v}_\textrm{3D} - {\bf v}_\textrm{cluster}
 \label{eq:9}
 \end{equation}
 where the caret denotes that this is the reconstructed velocity, rather than the measured velocity which is also available for a subset of the sample (see Fig.~\ref{fig:rvcomp}).  We do not account for the measurement errors in the parallax, position and proper motion in this analysis.   The typical relative uncertainties on these quantities is less than a few percent for the stars in the sample. For the cluster members themselves, the median distance error is 1.17~pc and the median proper motion error in one-dimension is 0.05 and 0.08 milliarcseconds per year in declination and right ascension respectively.  To place the proper motion error in context, this is $32-51$~m/s at the distance of the Pleiades cluster, and over the course of 130~Myr, this would result in a position error of $4-7$~pc.  These two distance errors are well within the threshold value of $d_\textrm{min}$ that we will use to identify potential escapees (15~pc).

\begin{figure}
    \centering
    \includegraphics[width=\columnwidth]{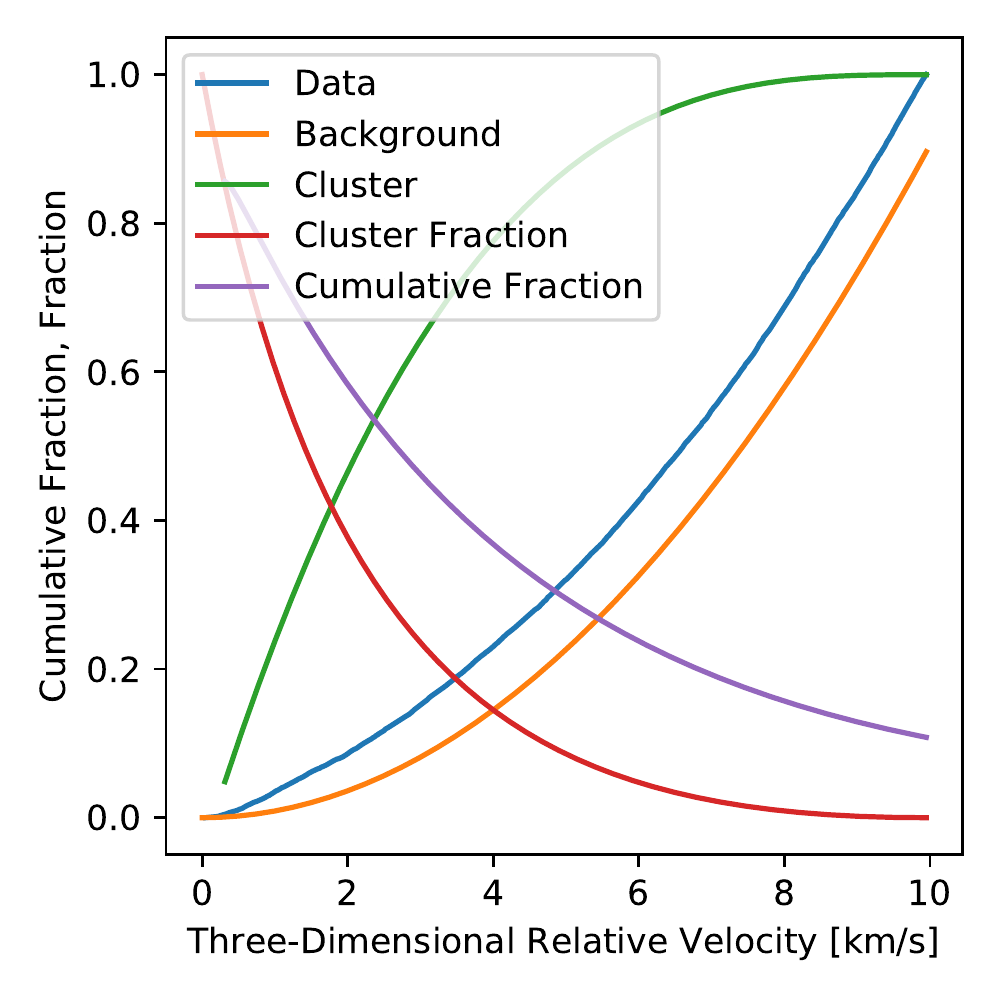}

    \caption{The cumulative distribution of reconstructed velocities for the stars that have come within 15~pc of the Pleiades in the past (blue).  There is an excess of stars at small velocities above the background (orange) which we model as uniform in two dimensions. The cumulative distribution of the excess over the background, the cluster, is shown in green.  The fraction of stars at a given velocity that are cluster members is given in red, and the cumulative fraction within a given velocity is depicted in purple. }
    \label{fig:frac3d}
\end{figure}

 To determine the threshold for the relative velocity to be deemed a candidate escapee, we look at the cumulative distribution of the magnitude of the three-dimensional reconstructed relative velocity $\Delta \hat {\bf v}_\textrm{3D}$ as shown by the blue curve in Fig.~\ref{fig:frac3d}.  We have restricted the sample to relative velocities less than 10~km/s and those stars that were less than 15~pc from the centre of the cluster in the past 140~Myr which we have taken as an upper limit on the age of the cluster \citep{2018A&A...616A..10G}.   With these constraints in mind, we can see that the magnitude of the reconstructed radial $v_r$ is strongly correlated with the magnitude of $\Delta {\bf v}$, so we expect the relative velocity of background stars to be uniformly distributed in two dimensions as shown by the orange curve which is parallel to the blue curve for large relative velocities.  With the background normalisation at large relative velocity, we can determine the cumulative fraction of cluster stars which we take to be the excess over the background (shown in green).  This yields the probability that a star with a given relative velocity is a member of the cluster distribution (red) and the fraction of cluster stars in a sample defined by a given maximum relative velocity (purple).

In particular we find that the distribution of the relative velocity of the escapee candidates is strongly peaked toward small velocities as expected from the relaxation of the cluster velocity distribution toward a Maxwellian distribution through two-body interactions The velocities of evaporated stars are typically $\sim\sqrt{2}v_\sigma$, where $v_\sigma$ is the cluster's velocity dispersion \citep{2008gady.book.....B}.  From the Gaia EDR3 proper motions, we estimate the three-dimensional velocity dispersion to be about 2.4~km/s, so we expect evaporated stars move up to about 3.4~km/s relative to the cluster.  Fig.~\ref{fig:frac3d} shows the cumulative distribution of cluster escapees in green that we measure over the background distribution in orange.  Although there are escapees up to and beyond 3.4~km/s, at a relative velocity of 2.36~km/s (approximately the cluster velocity dispersion), we find that both the cumulative fraction of cluster stars and the fraction of cluster stars within this relative velocity is 53.5\%.  We choose this as our threshold for a star to be an escapee candidate as at this value the true-positive rate equals the completeness rate (cumulative fraction of cluster stars) so that the total number of stars within the escapee candidate sample is expected to equal the total number of escapees. 

Fortunately, for a sub-sample of our escapee candidates, Gaia DR2 has measured radial velocities so that we can compare the reconstructed radial velocity $v_r$ (Eq.~\ref{eq:6}) with the measured value.  Fig.~\ref{fig:rvcomp} compares the reconstructed radial velocities with those observed for stars that we estimate to have left the cluster within the last fifty million years.  There is only a single star that we think left the cluster earlier with a Gaia DR2 radial velocity.  For the stars that have escaped within the last twenty-five million years there is excellent agreement between the reconstructed and observed radial velocities, with only a single outlier; therefore, we argue that the true positive rate for the recent escapers is likely to be larger than what inferred from Fig.~\ref{fig:frac3d}. For the stars that left earlier, there are more outliers, perhaps seven out of sixteen, meaning that 56\% of the escapees identified in this sample have reconstructed and observed radial velocities in agreement, similar to the true-positive rate shown in Fig.~\ref{fig:frac3d} of 53.5\%.

\section{Results}
\label{sec:results}

Having defined our sample of candidate escapees, we discuss the sample in further detail including their positions in \S~\ref{sec:positions}, the colour-magnitude diagram in \S~\ref{sec:escapee-cmd} and the effects of evaporation on the cluster in \S~\ref{sec:evaporation}.  We devote a separate section to look at the escaped white dwarfs (\S~\ref{sec:white-dwarfs}).

\subsection{Positions}
\label{sec:positions}

\begin{figure*}
    \centering
    \includegraphics[width=0.32\textwidth,trim=0 0.2in 0 0]{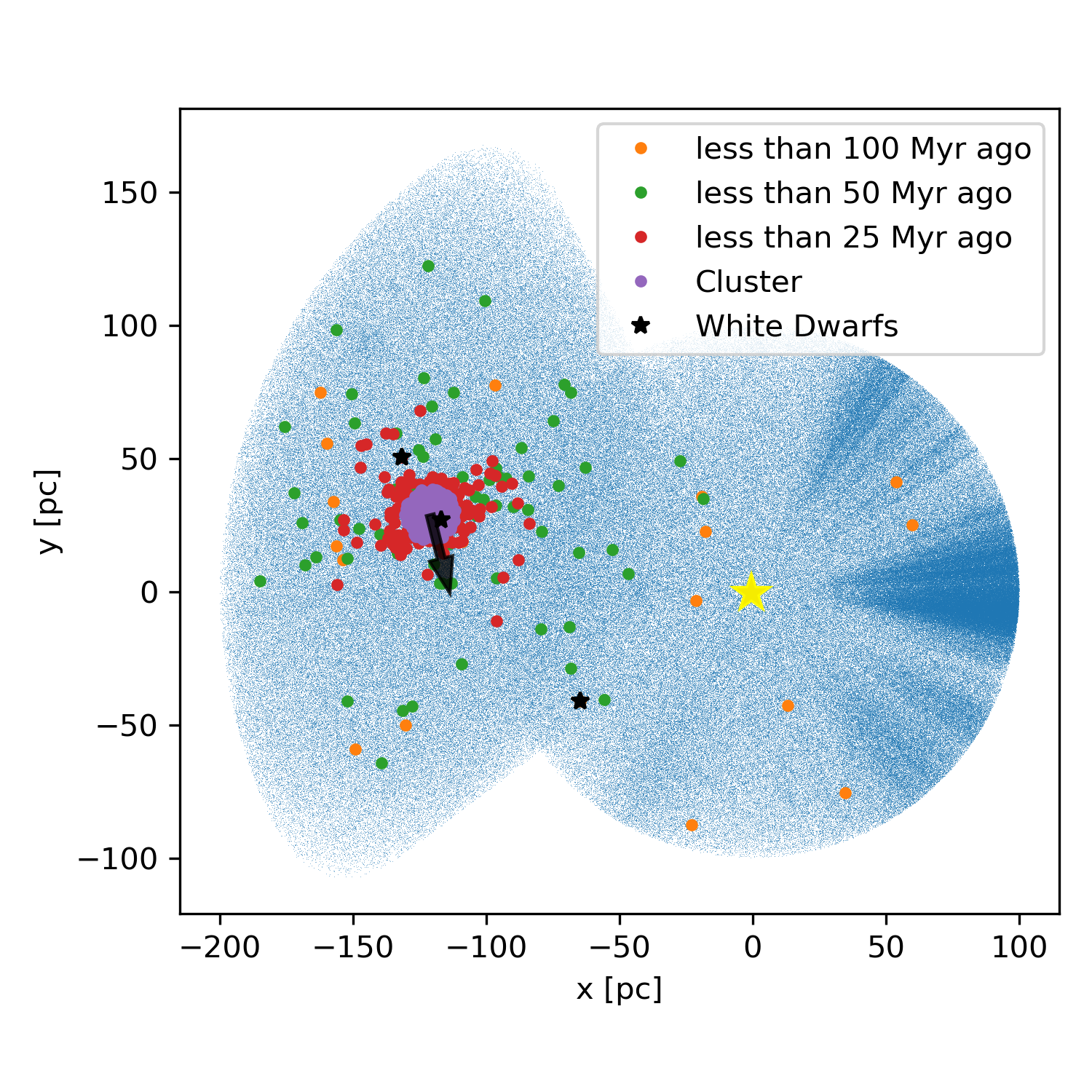}
    \includegraphics[width=0.32\textwidth,trim=0 0.2in 0 0]{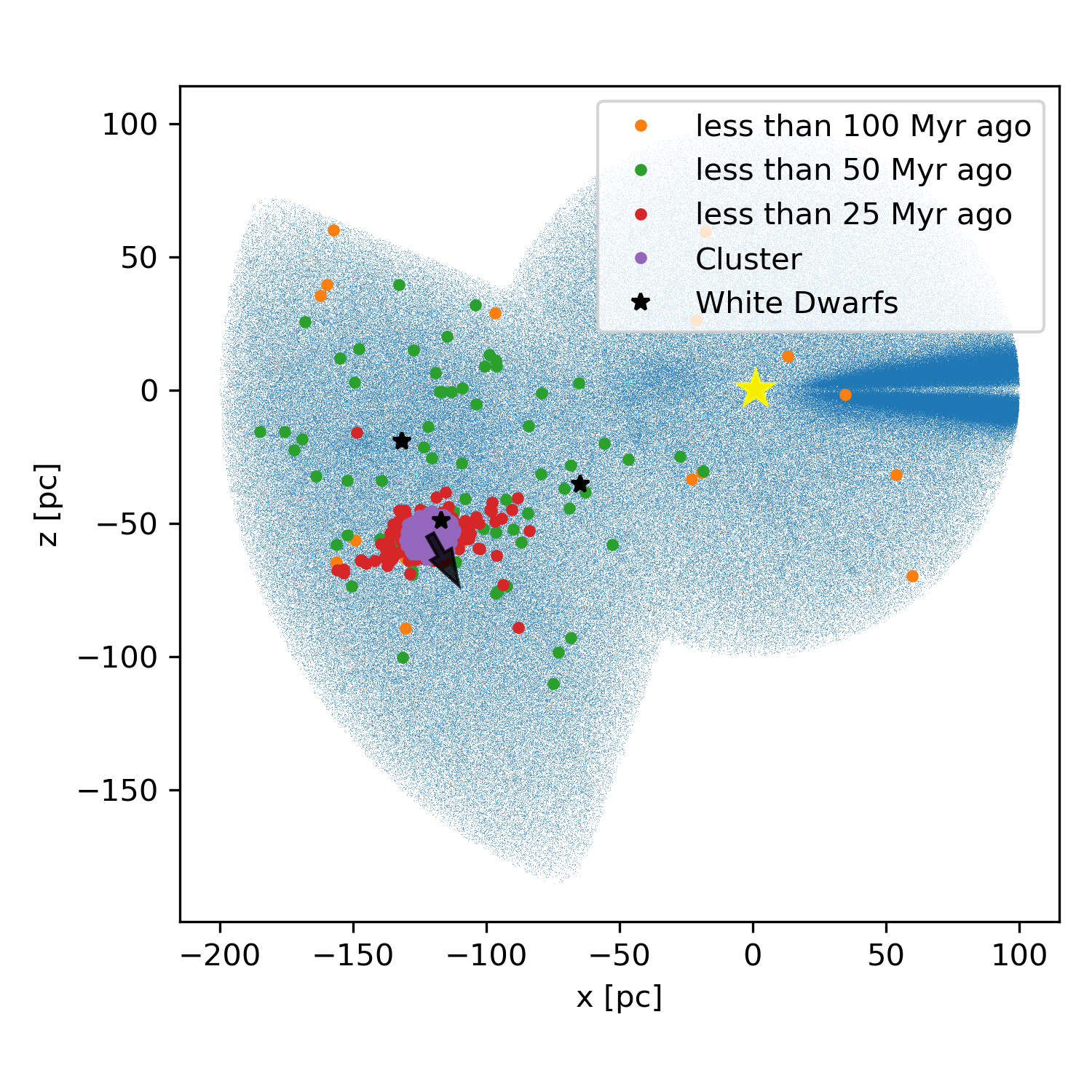}
    \includegraphics[width=0.32\textwidth]{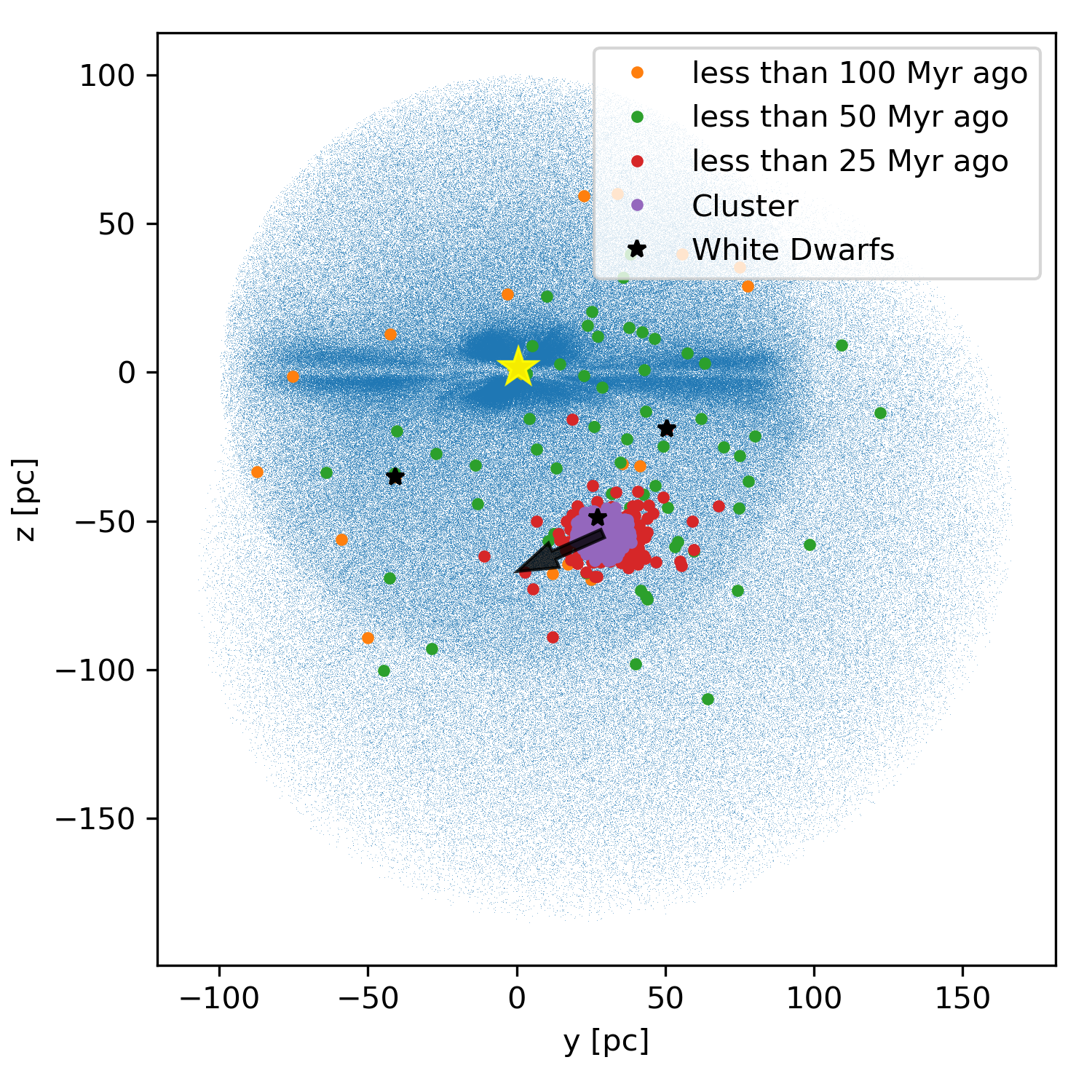}
        \caption{Where are they now?  The location of the Sun is depicted with a yellow star, and the displacement of the Pleiades over one million years relative to the local standard of rest is shown by the arrow. }
    \label{fig:esc-positions}
\end{figure*}
Fig.~\ref{fig:esc-positions} shows the positions of the candidate escapees, grouped as a function of the time of escape, with the cluster stars and the three white dwarfs identified as well.  The location of the Sun is given by a yellow star.  Even if we account for stars that may have left the sampling volume, it is apparent that the rate of evaporation from the cluster is increasing.  We have also indicated the motion of the cluster relative to the Galaxy with a black arrow.  To obtain the velocity of the cluster relative to the local standard of rest (LSR), we add the solar velocity relative to the LSR 
 \begin{equation}
 {\bf v}_\odot = \left (11.1^{+0.69}_{-0.75}, 12.24^{+0.47}_{-0.47}, 7.25^{+0.37}_{-0.36} \right )~\textrm{km/s}
 \end{equation}
to that of the cluster \citep{2010MNRAS.403.1829S}.  Apparently, the cluster passed through the Galactic plane a few million years ago, and perhaps this accounts for the recent increase in the rate of evaporation.

\begin{figure*}
    \centering
    \includegraphics[width=0.32\textwidth,trim=0 0.2in 0 0]{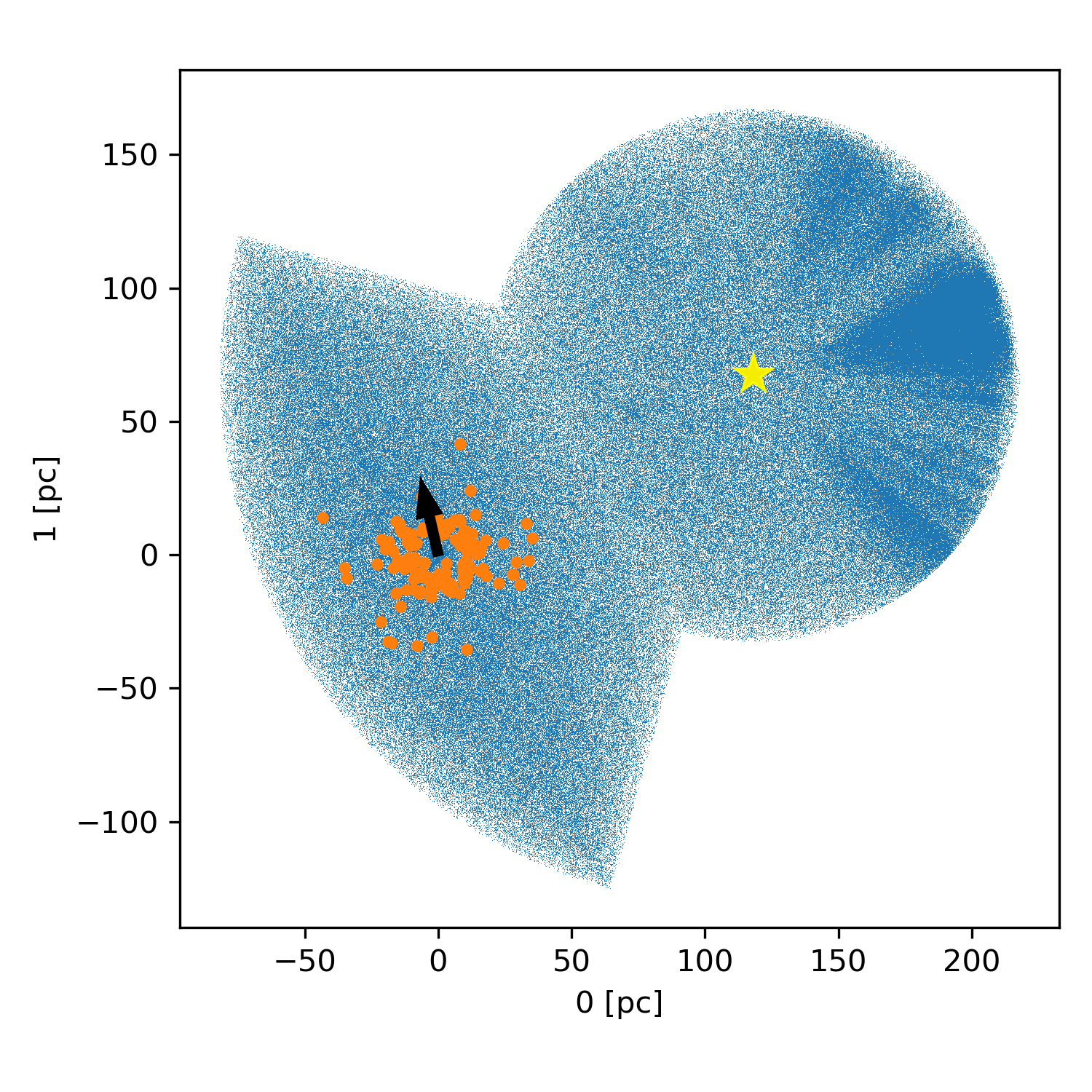}
    \includegraphics[width=0.32\textwidth,trim=0 0.2in 0 0]{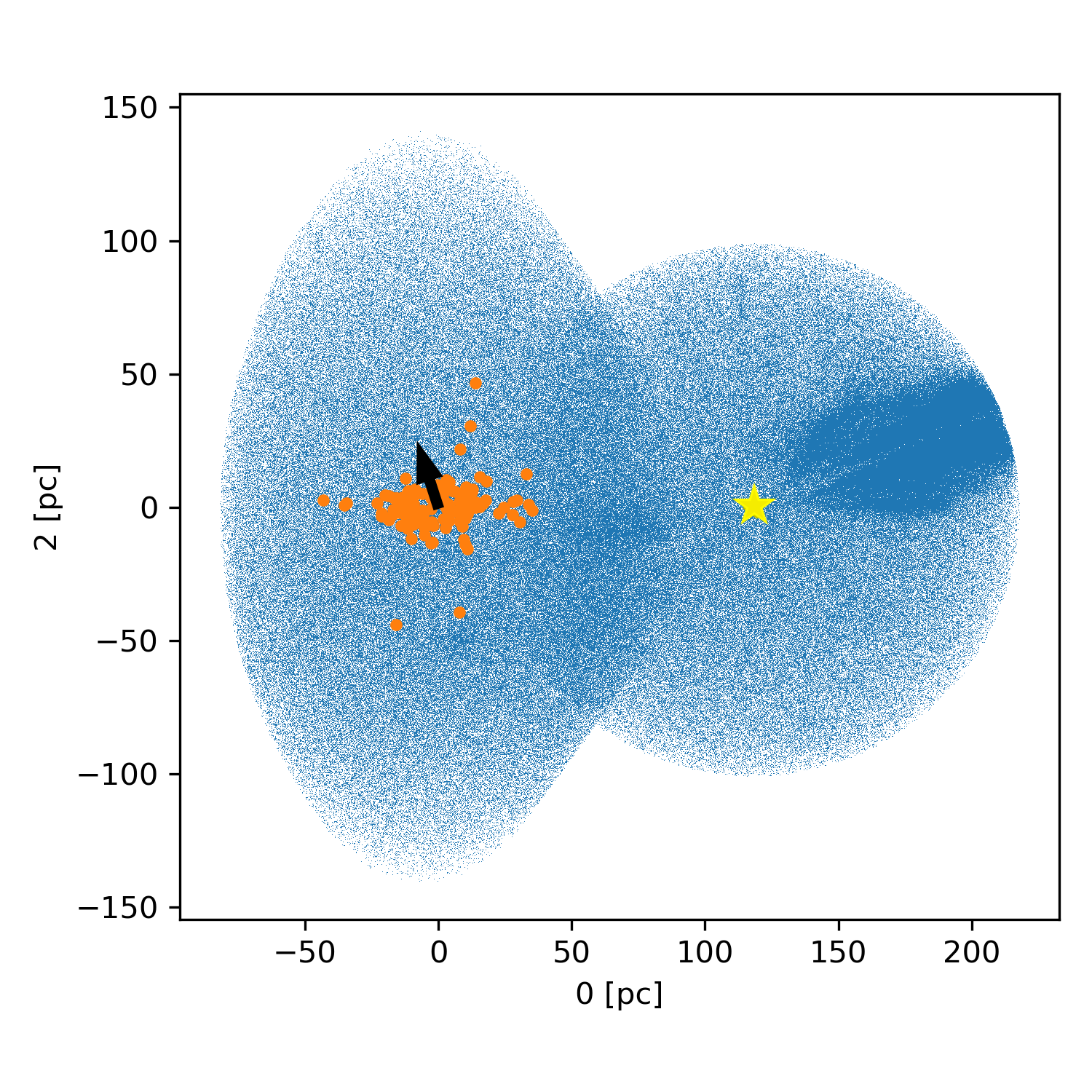}
    \includegraphics[width=0.32\textwidth,trim=0 0.14in 0 0]{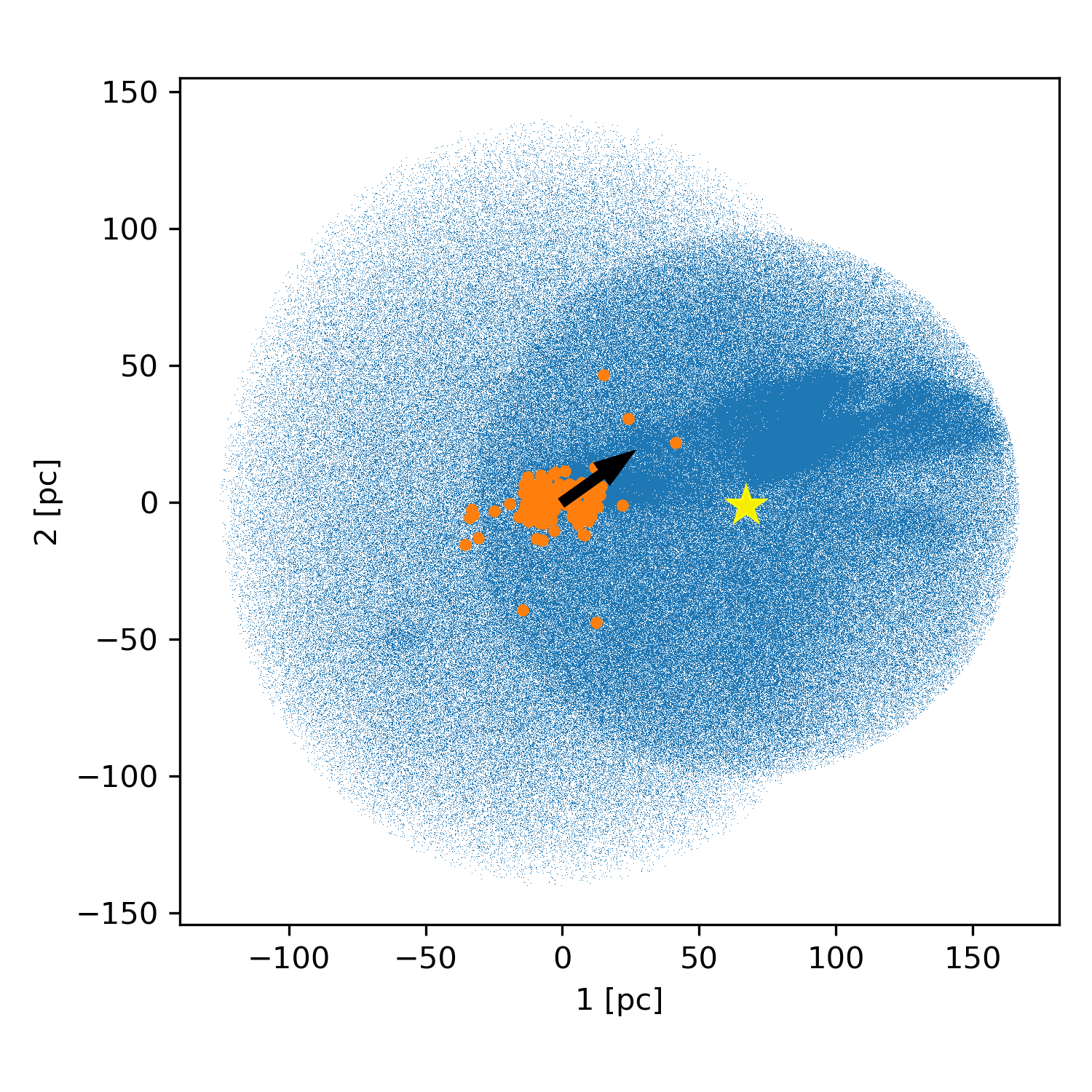}
        \caption{The location of the  25-Myr escapee sample relative to its principal axes.}
    \label{fig:eigen}
\end{figure*}

To understand better the anisotropy of the recently escaped stars we determine the body frame of those stars that have escaped within the last 25 Myr and are still within 30~pc of the centre of the cluster to reduce the effects of outliers. Fig.~\ref{fig:eigen} shows the positions of the escapees in the body frame of the distribution.  The axes are numbered (0 to 2) from the one with the smallest to the largest moment of inertia.  In Galactic coordinates, the three principal axes of the distribution are 
\begin{equation}
\left [
\begin{array}{ccc}
     X \\ 
     Y \\  
     Z
\end{array}
\right ] =
\left [
\begin{array}{rrr}
     0.90 \\ 
     0.29 \\  
     0.32
\end{array}
\right ],
\left [
\begin{array}{rrr}     
     0.22 \\ 
     -0.95 \\ 
     0.23 
\end{array}
\right ]
\textrm{and}
\left [
\begin{array}{rrr}     
     0.37 \\ 
     -0.14 \\ 
     -0.92
\end{array}
\right ].
\end{equation}
and the corresponding root-mean-squared radii are 11, 8 and 5~pc, so the distribution is closer to oblate but triaxial.  The minor axis is approximately perpendicular to the Galactic plane, and the major axis points approximately toward the Galactic centre.   The axes do not coincide closely with the velocity of the cluster or the displacement from the Sun.  The structure of the escapee population may have as much to do with the Galactic potential as the internal dynamics of the cluster as argued by \citet{2021A&A...645A..84M}. 

The structure of the population of candidate escapees from the past 25~Myr resembles strongly both the distribution of stars beyond the tidal radius identified by \citet{2019A&A...624A.110F} and the corona of the Pleiades found by \citet{2021A&A...645A..84M}.  Both these previous works looked at the spatial coincidence and contiguity of the stars with the cluster and proper motions similar to that of the bulk of the cluster.  Our method only requires that the stars with their measured velocities could have been close to the centre of the cluster in the past and have a small relative velocity to the cluster now.  Consequently, we find many objects not found by the previous searches, and we can extend our search to stars that left the cluster earlier so that the spatial coincidence is long lost.

The maximum relative velocity that a star can have and still be identified as a candidate escapee is 2.36~km/s so over the course of 25~Myr an escapee can only travel 60~pc from the centre of the cluster; therefore, all of the stars that have escaped from the cluster with velocities this small must still be in the sampled volume.  To estimate the fraction of the escapees that remain in the sampling volume after longer periods, we can simply evolve the reconstructed velocities of those stars that have escaped within the last 25~Myr to simulate the older samples.  Table~\ref{tab:volume} shows the number of stars that remain within the total volume ($N_1$) and the nearside volume ($N_{1/2}$) as time progresses.  We see that the nearside volume contains a larger fraction of the escapees than the total volume.  From the geometry in Fig.~\ref{fig:positions}, this result is expected.  However, because the distribution of the velocities of the escapees is anisotropic a larger number of stars is contained with the sampled volume than would be obtained if the velocity distribution were spherically symmetric.  The final columns give the fraction of escapees that remain in the volume and in particular we will use the second to last column $C_1$ to correct the samples of escapees to estimate the mass function of the cluster as a function of time and the total mass of stars that have left the cluster.
\begin{table}
     \caption{Volumetric Completeness: The subscript 1 refers to the total volume, and the $1/2$ refers the nearside volume shown in orange in Fig.~\ref{fig:positions}.}
     \centering
     \begin{tabular}{c|rrrrr}
     \hline
     Escape Time & $d_\textrm{max}$ & $N_1$ & $N_{1/2}$  & $C_1$ & $C_{1/2}$ \\
     
     [Myr] & [pc] \\
     \hline
$0-25$ & 60 & 126 & 57 & 1.00 & 1.00 \\
$25-50$ & 121 & 116 & 56 & 0.92 & 0.98 \\
$50-75$ & 181 & 93 & 54 & 0.73  &  0.95 \\
$75-100$ & 242 & 76 & 48 & 0.60 & 0.84 \\
$100-125$ & 302 & 58 & 37 & 0.46 & 0.65 
     \end{tabular}
     \label{tab:volume}
 \end{table}

\subsection{Colour-Magnitude Diagram}
\label{sec:escapee-cmd}

\begin{figure}
    \centering
    \includegraphics[width=\columnwidth]{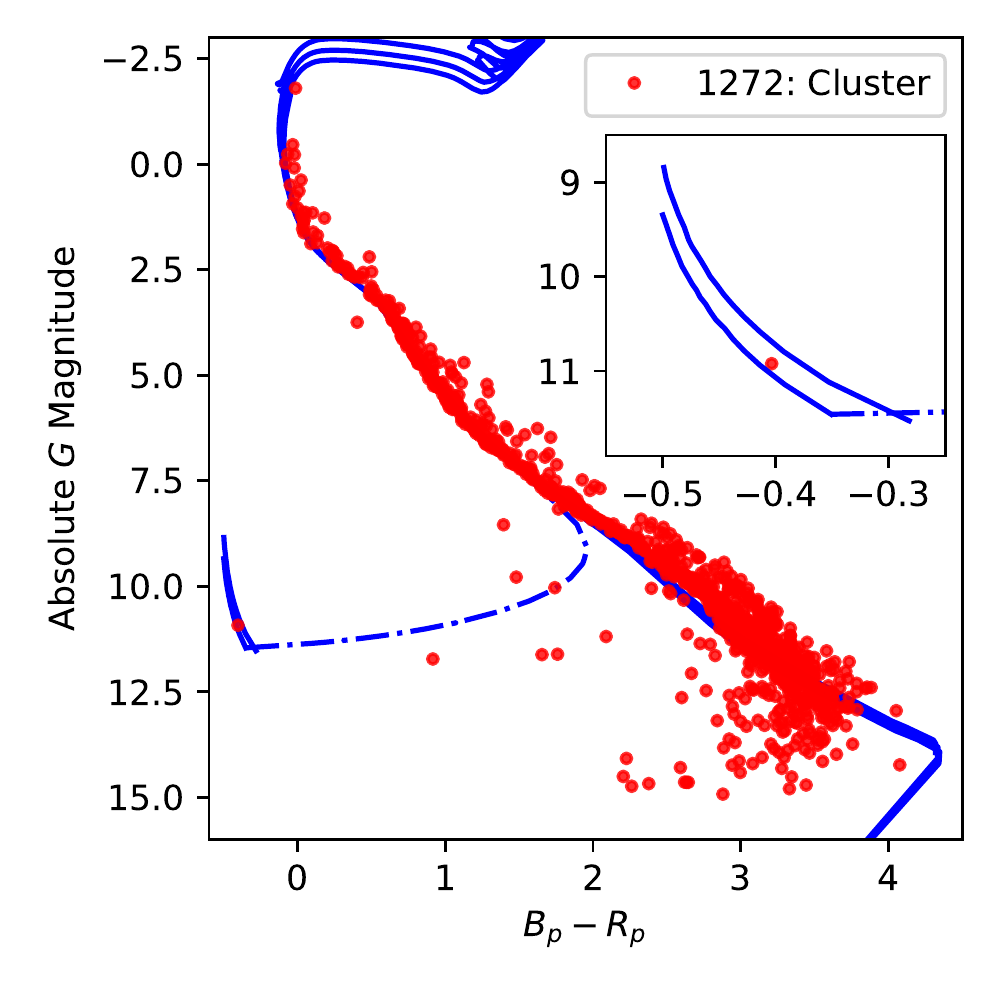}
    \caption{Colour-magnitude diagram of the Pleiades today using the volume-limited sample of those objects within 10~pc of the centre of the cluster and with proper motions lying within the range shown in Fig.~\ref{fig:pmselection}.  The blue curves are as in Fig.~\ref{fig:cmd=pmselection}.}
    \label{fig:cmd}
\end{figure}
We will first focus on the colour-magnitude diagram (CMD) of the cluster stars that we define here to be those stars within 10~pc of the centre of the cluster in three dimensions  \citep[similar to the tidal radius of the cluster][]{2019A&A...628A..66L,2021A&A...645A..84M} and also have proper motions within five milliarcseconds per year of the median cluster proper motion.  The CMD of this sample is depicted in Fig.~\ref{fig:cmd}.  This group of nearly 1,300 stars is slightly different from the sample depicted in Fig.~\ref{fig:cmd=pmselection} which had no constraint on the distance to the star from the Sun but a stronger constraint on the distance on the sky from the centre of the cluster.  We identify fewer stars above the turnoff in this sample, indicating that a few of the turn-off stars in Fig.~\ref{fig:cmd=pmselection} had measured parallaxes that placed them beyond the ten-parsec cutoff.  This is not surprising as the parallax uncertainties increase for the brightest stars in the sample, so these stars are likely to lie within the core of the cluster, but the uncertain measurements place them in front or behind.  This more broad sample across the sky includes the Pleiades white dwarf EGGR~25.  

\begin{figure*}
    \centering
    \includegraphics[width=0.32\textwidth]{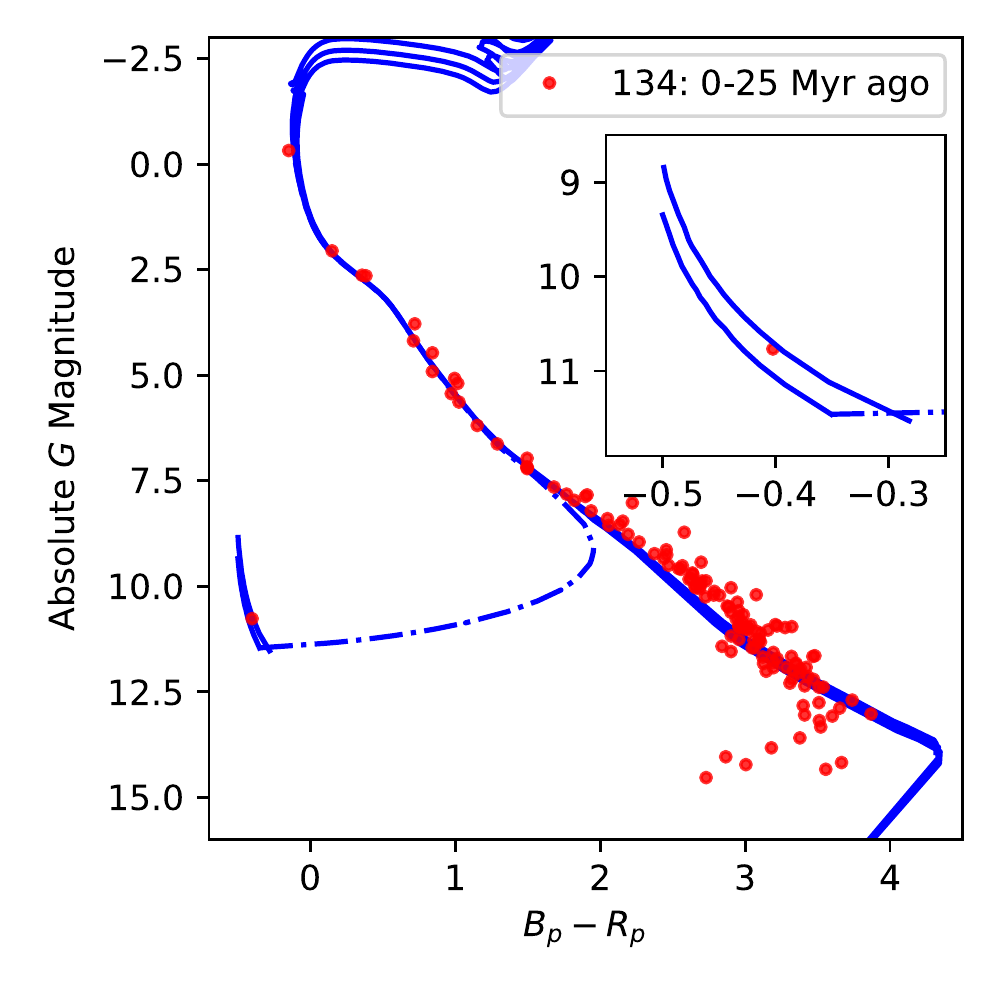}
    \includegraphics[width=0.32\textwidth]{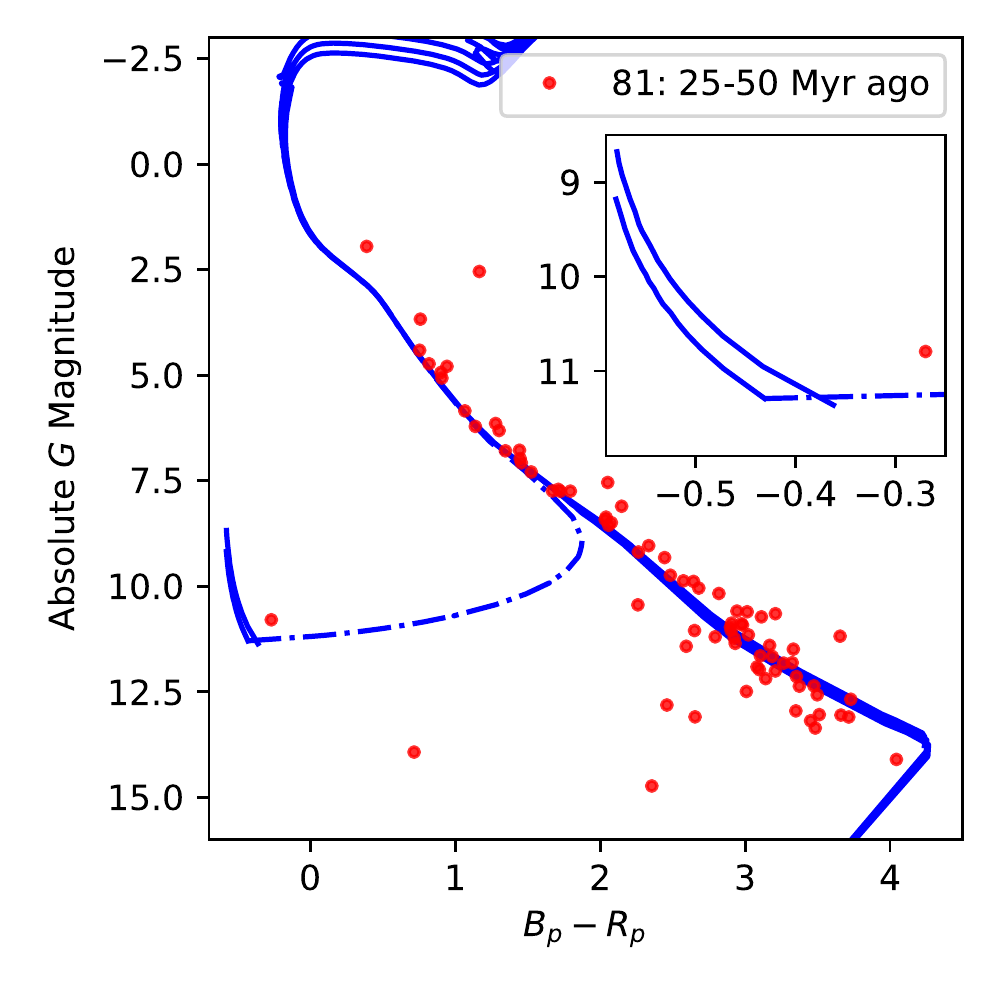}
    \includegraphics[width=0.32\textwidth]{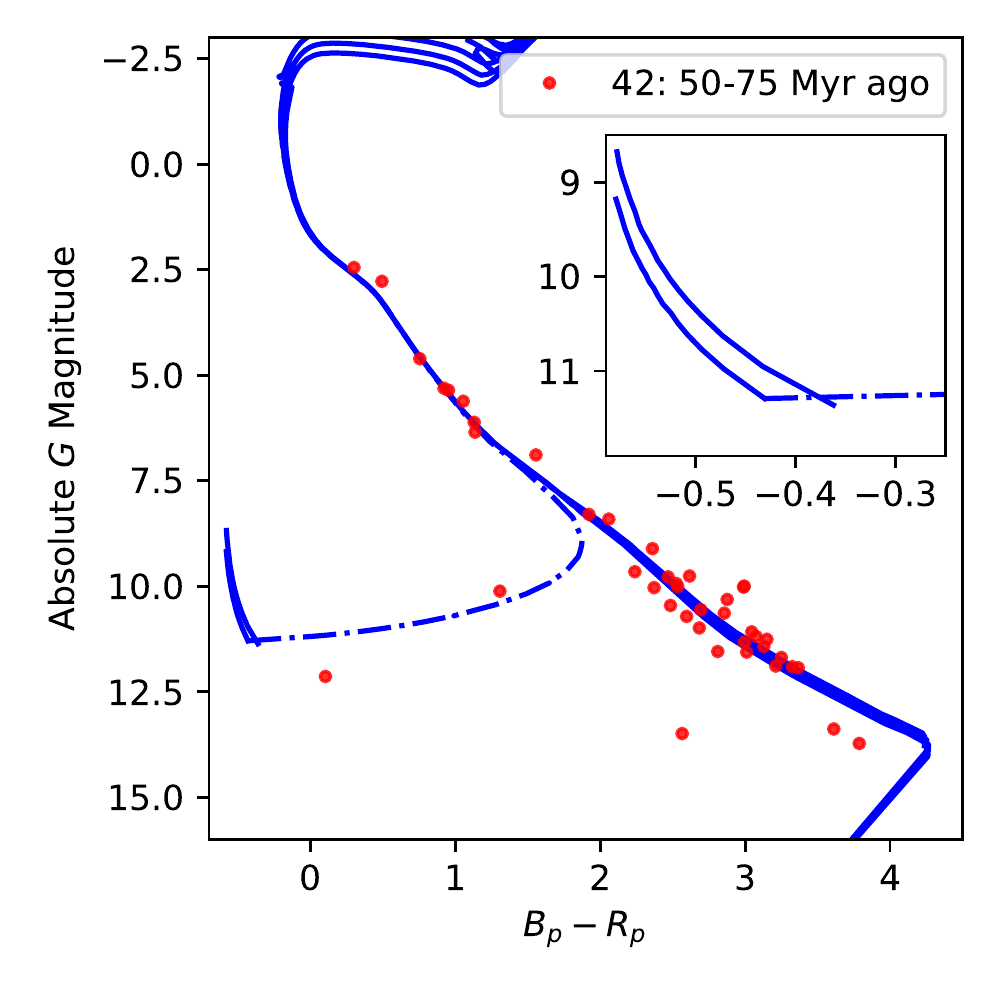}
    \includegraphics[width=0.32\textwidth]{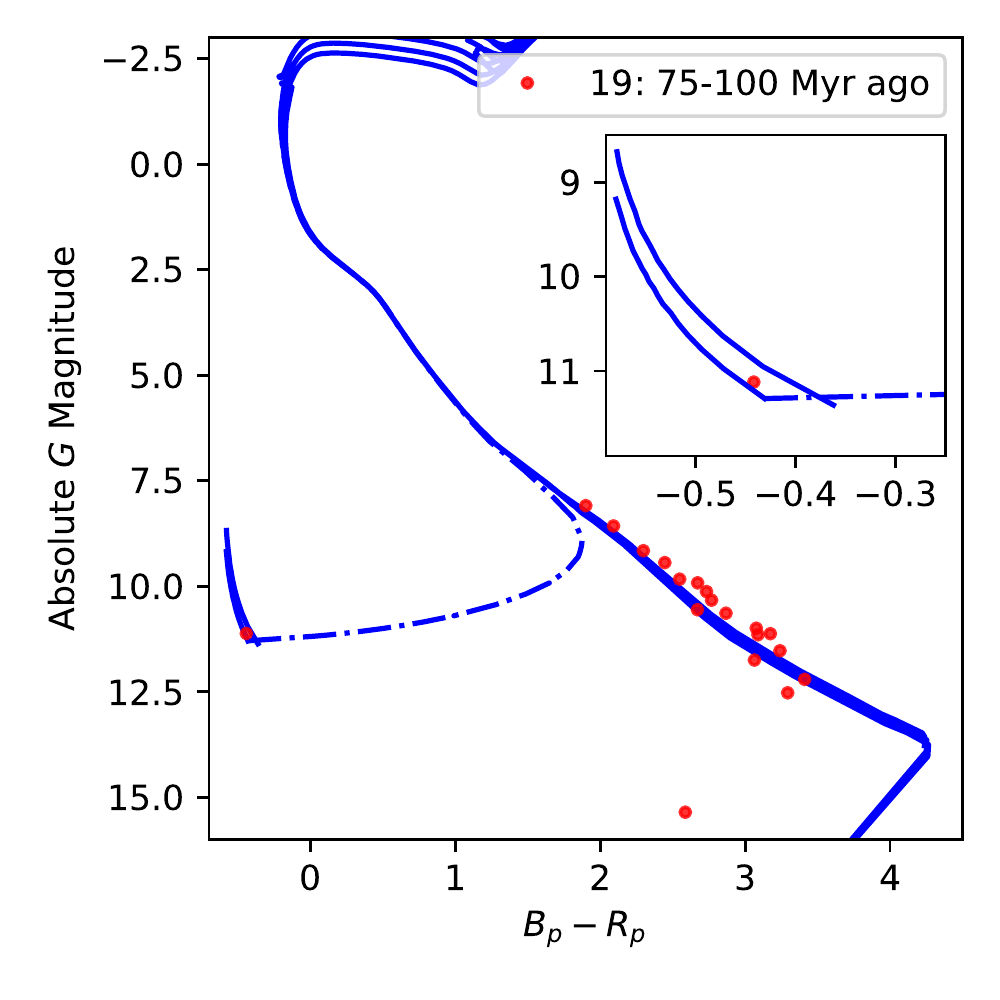}
    \includegraphics[width=0.32\textwidth]{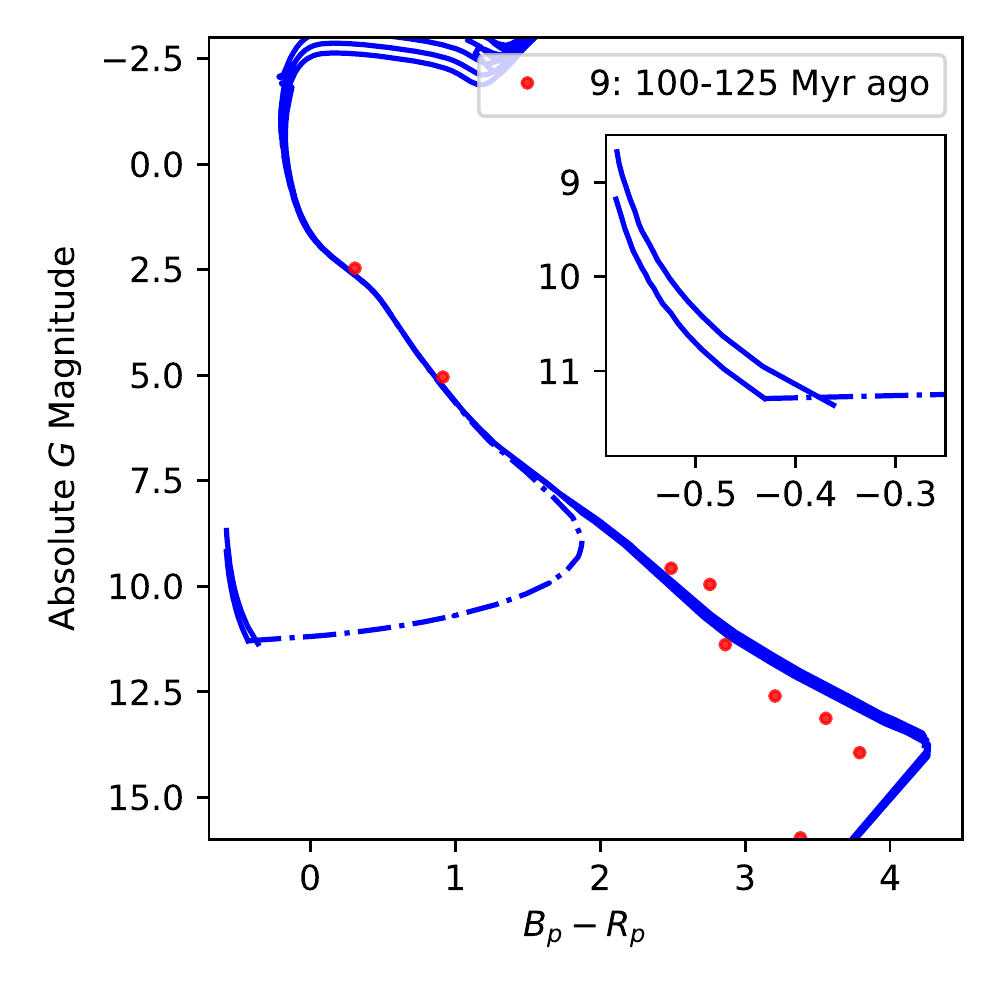}
    \includegraphics[width=0.32\textwidth]{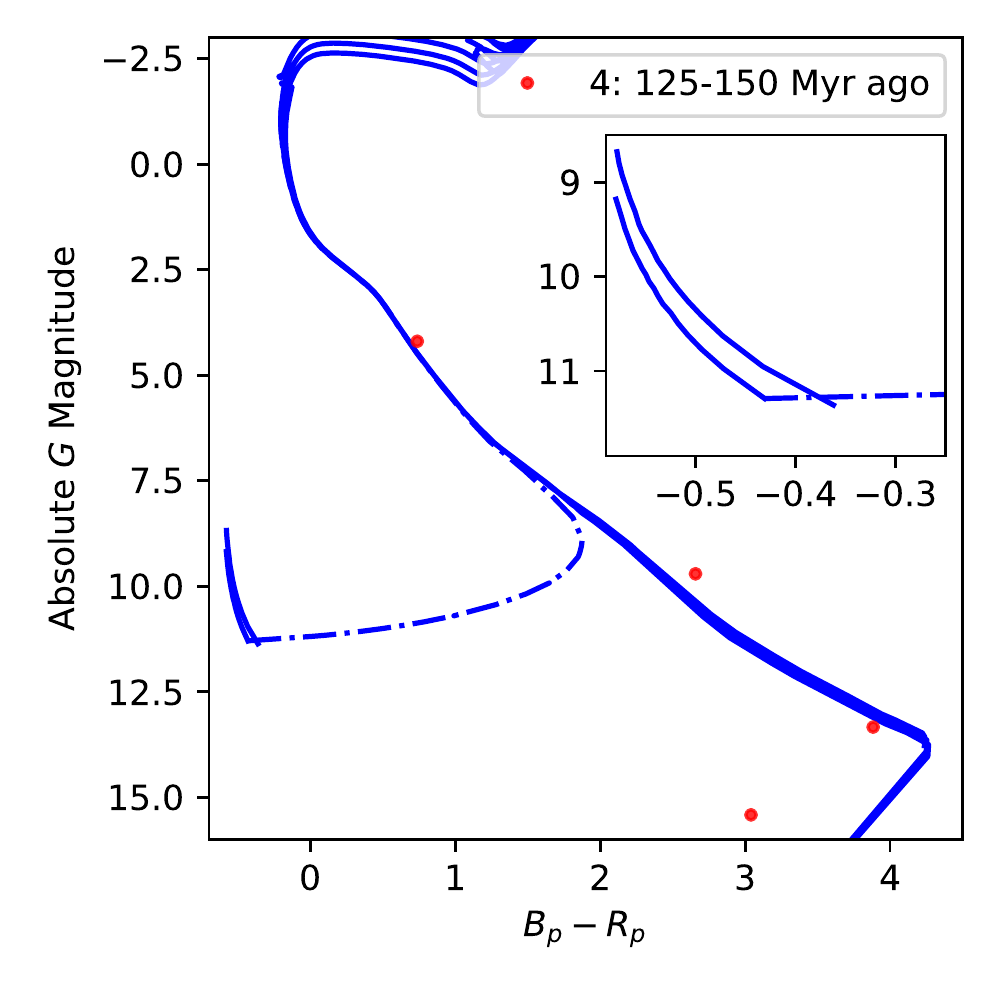}
    \caption{Colour-magnitude diagram of the Pleiad escapees as a function of time.  The blue curves are as in Fig.~\ref{fig:cmd=pmselection} with the cluster extinction applied for the 25~Myr sample and no extinction applied for the older samples.  The individual parallaxes of the objects are used to calculate the absolute magnitudes.}
    \label{fig:cmd-escape}
\end{figure*}

The stellar population of the Pleiades itself is well known, so we now proceed to examine the candidate escapees in Fig.~\ref{fig:cmd-escape}.  We first focus on the most recent escapees.  These stars, as we saw in \S~\ref{sec:positions}, lie close to the cluster in a triaxial configuration more than 10~pc from the centre of the cluster. Because the long axis of the configuration lies approximately along the line of sight, many of the stars in the 25-Myr escapee sample have already been listed as cluster members, but in fact the Gaia EDR3 parallaxes place them either closer or further from the cluster.  The colour-magnitude diagram of the most recent escapees resembles that of the cluster but it is necessarily more sparse.  As we look at stars that escaped the cluster earlier in time, the distribution is more and more skewed toward lower-mass stars, which is expected from mass segregation within the cluster \citep{1997A&ARv...8....1M,1998MNRAS.295..691B,1998A&A...331...81P}

The brightest candidate escapee both in an absolute and apparent sense is the fifth-magnitude star 41 Tau.  In the colour-magnitude diagram at the top left of Fig.~\ref{fig:cmd-escape}, it lies on a younger isochrone than the cluster itself; however, the isochrone has the extinction of the cluster applied.  41 Tau lies about five degrees away from the centre of the cluster and 14~pc closer, so one would expect somewhat less extinction than for the cluster itself.   The most recent published radial velocity for 41~Tau is $2.3 \pm 4.2$~km/s  \citep{2006AstL...32..759G} in agreement with the reconstructed values of 3.4~km/s.

Other than the most luminous member of the escapee population, several others stand out.  The nearest Pleiades escapee candidate to the Sun at 34~pc is the low-mass star LP~261-75.  LP~261-75 has already been identified as moving with the Pleiades and is very well studied as it hosts a twenty-Jupiter-mass planet.  In fact, it is the only object in the escapee sample that is suspected to host a planetary system. However, \citet{2012ApJ...758...56S} measured a radial velocity for this object of $10.2\pm 0.2$~km/s, in disagreement with the reconstructed value of $-4.1$~km/s, so although tantalising, this object is unlikely to be associated with the Pleiades.  If we continue to look at the nearest twenty-five objects in the sample, seven have measured radial velocities. All seven of these objects with radial velocity measurements are currently considered high probability members of the AB~Dor moving group.  And these seven objects are the only objects in the candidate escapee sample that are among the 760 members of the AB Dor moving group within 80~pc.  Fewer than one percent of nearby AB Dor members have proper motions consistent with being fewer than 15~pc from the centre of the Pleiades in the past with relative velocities less than 2.36~km/s.  Of these seven measurements, five agree with the reconstructed radial velocity, consistent with the rate expected from Fig.~\ref{fig:frac3d}. This result is not independent of the previous test as four of the seven objects have Gaia DR2 radial velocity measurements. The objects whose reconstructed radial velocities are in agreement with the measured ones left the cluster between 45 and 110~Myr ago. 
This group includes two RS Canum Venaticorum variables, V493~And and TY~Col. Several stars apparently escaped from the Pleiades very early in its lifetime. TY Col escaped from the cluster 110~Myr ago.  It is the second brightest star to escape between 100 and 125~Myr ago. The brightest star to escape during this epoch has already been associated with the Pleiades as a common-proper motion object. The twenty-fifth closest candidate escapee is the white dwarf WD~0518-105 at 84~pc. These long-lost sisters of the Pleiades typically have very small three-dimensional velocities relative to the cluster, increasing the \textit{a priori} likelihood that they are indeed former cluster members (see Fig.\ref{fig:frac3d}).

\subsection{Evaporation}
\label{sec:evaporation}

\begin{figure}
    \centering
    \includegraphics[width=\columnwidth]{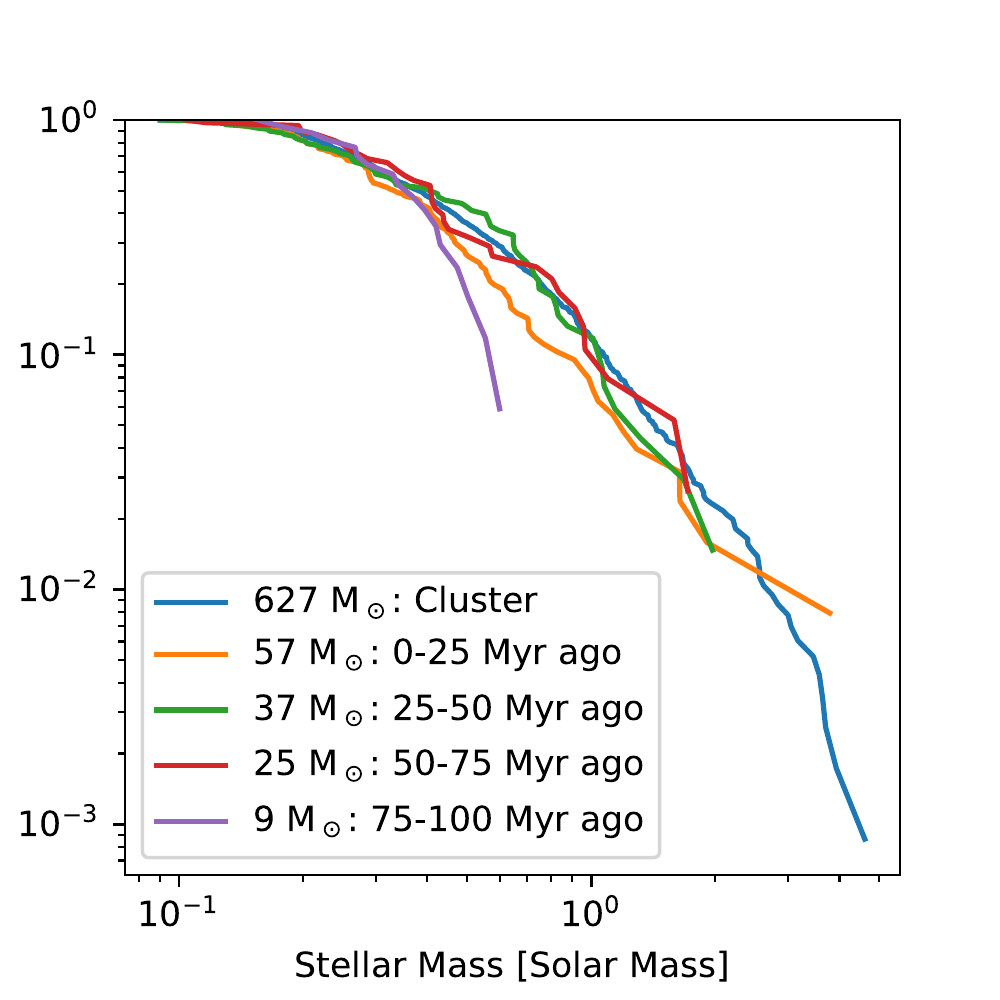}
    \caption{Cumulative Distribution of Escaped Stars as a Function of Mass}
    \label{fig:escape-mass}
\end{figure}

We can use the escaped stars depicted in the colour-magnitude diagrams in Fig.~\ref{fig:cmd-escape} to estimate the mass function of the escaped stars and how the mass function of the cluster has evolved.  To do this, we focus on the stars that lie along the main sequence and use the 140-Myr Padova isochrone to infer the initial mass of each star.  The resulting cumulative mass distributions are depicted in Fig.~\ref{fig:escape-mass}.  The total mass given in the legend includes a correction for the escapees that have left the sample region (from Table~\ref{tab:volume}).  This correction is greater than a factor of two for stars that have left the cluster more than 100~Myr ago.  Because of this, and of the small numbers of early escapees, we only consider stars that have escaped within the last 100~Myr.  

\begin{figure}
    \centering
    \includegraphics[width=\columnwidth]{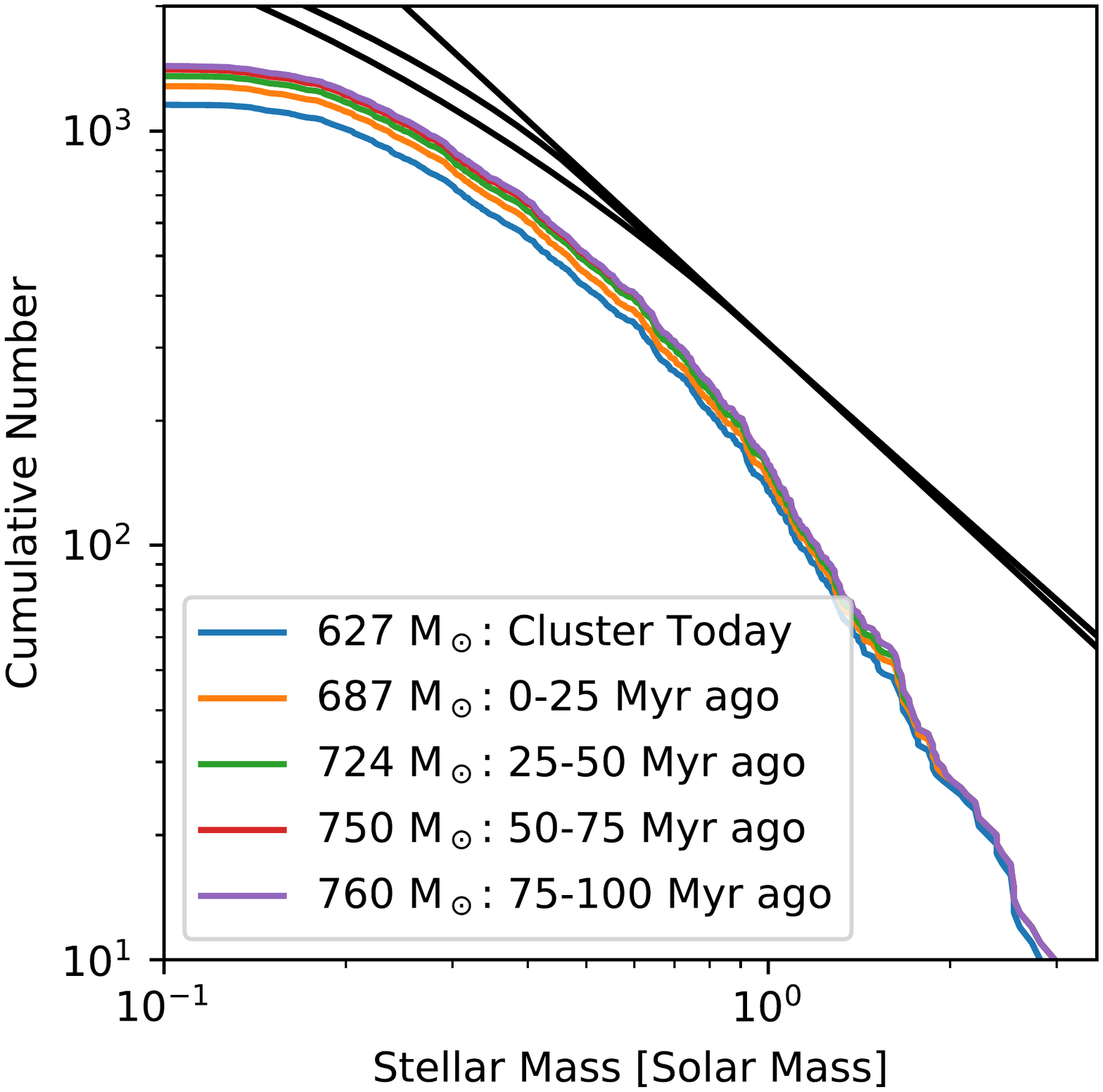}
    \caption{Cumulative mass function of the Pleiades today and in the past.  The black curves from top to bottom trace  \citet{1955ApJ...121..161S}, \citet{2001MNRAS.322..231K} and \citet{2003PASP..115..763C} initial-mass functions.}
    \label{fig:cum-mass-funk}
\end{figure}

Fig.~\ref{fig:cum-mass-funk} accumulates the mass lost from the cluster with the current mass function of the cluster to reveal how the mass function of the cluster has evolved.  The bulk of the stars lost have had masses much smaller than one solar mass.  Initially the cluster had more than 1,600 stars with a mean mass of 0.4~M$_\odot$.  Today the cluster has nearly 1,300 stars with a mean mass of 0.5~M$_\odot$.  The mean mass of the 300 lost stars is 0.42~M$_\odot$.  Although the mass loss rate has been increasing over the past 100~Myr, the mean mass of the lost stars has held steady, except in the earliest epoch of $75-100$~Myr when no massive stars left the cluster, yielding a mean mass of 0.3~M$_\odot$ for stars lost during this period.  If the white dwarf WD~0518-105 left the cluster before becoming a white dwarf, this would increase the mean mass during this epoch to 0.6~M$_\odot$, and in fact this object would have been more than twice as massive as 41~Tau, the most massive star in Fig.~\ref{fig:cmd-escape}; consequently, we argue that WD~0518-105 is more likely to have formed within the cluster or perhaps that its formation triggered its escape through a kick \citep{1998A&A...333..603S,2003ApJ...595L..53F,2008MNRAS.383L..20D,2007MNRAS.381L..70H,2008MNRAS.385..231H,2008MNRAS.390..622H}.  Three commonly used initial mass functions \citep[IMFS,][]{1955ApJ...121..161S,2001MNRAS.322..231K,2003PASP..115..763C} are depicted for illustration.  Although the slope at about 0.5~M$_\odot$ is similar between the data and the IMFs shown, the cluster has a deficit of both low-mass and high-mass stars relative to the IMFs.

\section{White Dwarfs}
\label{sec:white-dwarfs}

From an examination of Fig.~\ref{fig:cmd-escape} we can see several objects in the white-dwarf region.  However, only two objects lie on the cooling tracks for young and massive white dwarfs, the only ones that one would acceptable to be associated with the Pleiades (between the blue lines in the inset).  In fact, even if one relaxes the mass constraint, these are the only young white dwarfs in the escapee sample.  The astrometric and derived quantities for these objects are presented in Tables~\ref{tab:wd_photo} and~\ref{tab:wd_derive}.  The younger of the two white dwarfs, Lan 532, lies relatively close to the cluster and has a relative velocity of 2.11~km/s with respect to the cluster.  Although the \textit{prima facie} probability that one would assign for this object to be an escapee is about 35\%, the additional facts that young (less than 50~Myr) and massive white dwarfs (greater than one solar mass) are rare in the field \citep{massive} dramatically increases the level of confidence that Lan~532 left the Pleiades about 20~Myr ago and became a white dwarf about 50~Myr ago.   The second escaped white dwarf is WD~0518-105.  Its velocity relative to the cluster is 1.07~km/s, yielding a \textit{prima facie} probability of association of nearly 60\% from Fig.~\ref{fig:frac3d}.  The arguments about the rarity of young, massive white dwarfs apply equally strongly to this object. We conclude that it is much more likely that WD~0518-105 is also a long-lost Pleiad that both left the cluster and became a white dwarf about 60~Myr ago rather than a rare interloping young massive field white dwarf.
\begin{table*}
\caption{Pleiad White Dwarfs (Photometric, Spectroscopic and Astrometric Quantities)}
\label{tab:wd_photo}
\begin{tabular}{llrrrrrrrr}
\hline
       Gaia Source ID                   & Name & \multicolumn{1}{c}{RA}     &  \multicolumn{1}{c}{Dec}     & \multicolumn{1}{c}{Abs $G$}  &  \multicolumn{1}{c}{Parallax} & \multicolumn{1}{c}{$B_p-R_p$}  & \multicolumn{1}{c}{$T_\textrm{eff}$} & \multicolumn{1}{c}{$\log g$}
       \\
                                        &      & \multicolumn{1}{c}{[deg]}  &  \multicolumn{1}{c}{[deg]}   &  \multicolumn{1}{c}{[mag]}   &   \multicolumn{1}{c}{[mas]}   & \multicolumn{1}{c}{[mag]} & \multicolumn{1}{c}{[$10^3$ K]} & \multicolumn{1}{c}{[cm s$^{-2}$]}           \\
                            
\hline
Gaia EDR3 228579606900931968  & Lan 532     & 63.1792 & 40.6902    & 10.7666 &  7.030 & $-$0.4024 & $33.7\pm0.3$ & $8.61\pm0.05$ \\ 
Gaia EDR3 66697547870378368   & EGGR 25     & 58.0473 & 24.9300    & 11.0253 &  7.724 & $-$0.4035 & $33.0\pm0.2$ & $8.66\pm0.03$ \\
Gaia EDR3 3014049448078210304 & WD 0518-105 & 80.3291 & $-$10.4884 & 11.1176 & 11.870 & $-$0.4419 & $33.7\pm0.4$ & $8.70\pm0.06$ \\
\end{tabular}
\end{table*}
\begin{table*}
\caption{Pleiad White Dwarfs (Derived Quantities)}
\label{tab:wd_derive}
\begin{tabular}{lrrrrcccccc}
\hline
       Gaia Source ID                   & \multicolumn{1}{c}{$\Delta v_{2D}$} &  \multicolumn{1}{c}{$d_\textrm{present}$} & \multicolumn{1}{c}{$v_r$} &  \multicolumn{1}{c}{$d_\textrm{min}$} & $\Delta v_\textrm{3D}$ &  $t_\textrm{escape}$ & $v_g$ & Mass & $t_\textrm{cool}$  & Initial Mass \\
                                         &  \multicolumn{1}{c}{[km/s]}   &  \multicolumn{1}{c}{[pc]}                 & \multicolumn{1}{c}{[km/s]} & \multicolumn{1}{c}{[pc]} & [km/s] & [Myr] 
                                          &  \multicolumn{1}{c}{[km/s]} &
                                         [M$_\odot$] & [Myr] & [M$_\odot$] 
\\
\hline
Gaia EDR3 228579606900931968  & 2.11 & 42.64   & $-0.89$ & 10.31 & 2.22 & 18 & $78\pm28$ & $1.01\pm0.03$ & $35\pm6$
& $5.5^{+0.3}_{-0.3}$ \\
Gaia EDR3 66697547870378368   & 0.94 & 7.11    & 5.69  & $-$  & $-$ & $-$  &  $84\pm9$\phantom{0} & $1.04\pm0.02$ & $47\pm5$
& $5.8^{+0.3}_{-0.4}$ \\ 
Gaia EDR3 3014049448078210304 & 1.05 & 91.72  & 24.98 & 6.80 & 1.07 & 84 & $105\pm10$\phantom{1} & $1.06\pm0.04$ & 
$59\pm9$ 
& $6.2^{+0.4}_{-0.5}$ 
\end{tabular}
\end{table*}

\begin{figure*}
    \centering
    \includegraphics[width=\textwidth]{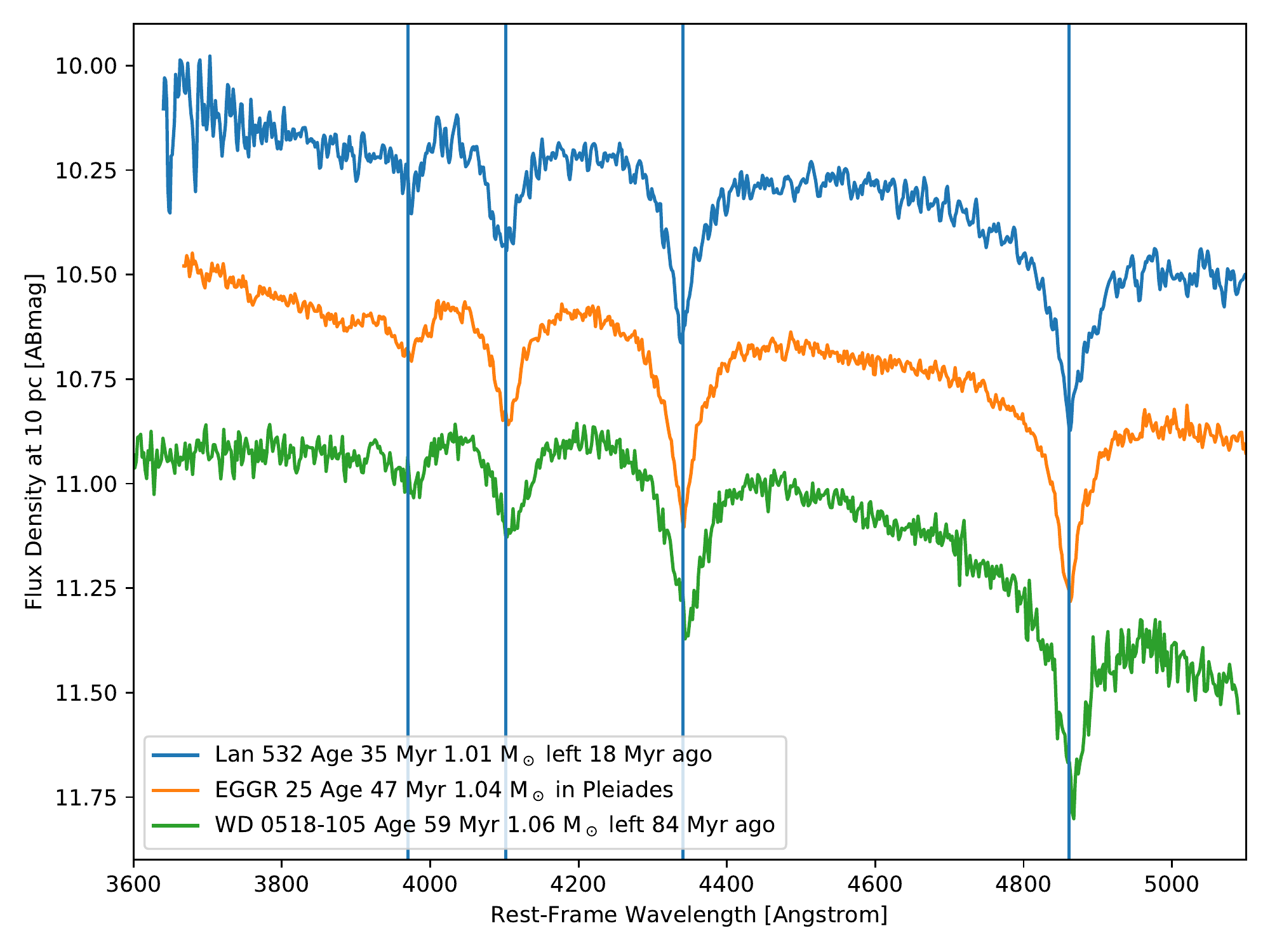}
    \caption{Spectra of the Three Pleiad White Dwarfs, from top to bottom: Lan 532, EGGR 25 and WD 0518-105. The spectra of EGGR 25 and WD~0518-105 are from \citet{2011ApJ...743..138G}.  The spectrum of Lan 532 is from \citet{2011AJ....141...96L}. The vertical lines indicate the rest-frame wavelengths of the hydrogen absorption lines of the Balmer series from H$\beta$ through H$\epsilon$ right to left.}
    \label{fig:Pleiad-WDs}
\end{figure*}

This yields a sample of three white dwarfs that have originated in the Pleiades when combined with the previously known EGGR 25.  Fig.~\ref{fig:Pleiad-WDs} displays the visual spectra of the objects in their rest frames using the inferred radial velocity, or the radial velocity of the cluster in the case of EGGR 25.  The flux density has been normalised to absolute AB magnitudes.  The progression from the youngest, brightest and least massive Lan~532 to the most massive WD~0518-105 is apparent.  The rest-frame wavelengths of Balmer-$\beta$ through Balmer-$\epsilon$ are indicated by the vertical lines.  

\begin{figure*}
    \centering
    \includegraphics[width=\textwidth]{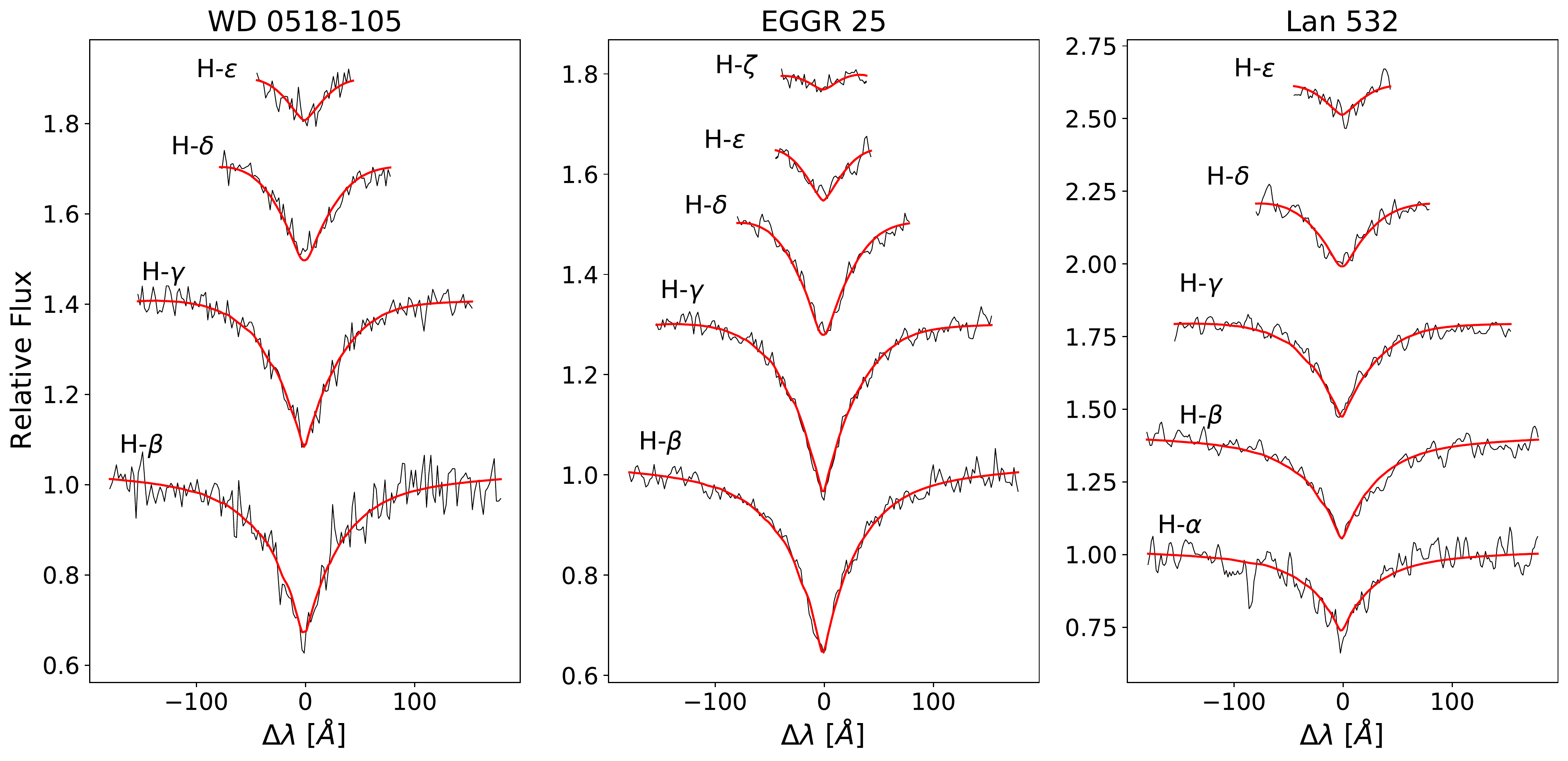}
    \caption{Hydrogen Balmer lines for the three Pleiad white dwarfs are displayed. The best fitting spectral models are superimposed.}
    \label{fig:Pleiad-Hlines}
\end{figure*}

We now focus on the three best detected Balmer lines as shown in Fig.~\ref{fig:Pleiad-Hlines} along with the best fitting white-dwarf atmosphere models.  The results of these atmospheric fits are given in Tab.~\ref{tab:wd_photo}. 
We also combine the redshift measurements (the sum of Doppler and gravitational redshift) of \citet{2020A&A...638A.131N} for WD~0518-105 of $130\pm10$~km/s with the reconstructed radial velocity to obtain a gravitational redshift for WD~0518-105 of $105\pm10$~km/s, slightly larger than the value obtained by \citet{1991ApJ...376..186W} for EGGR~25 of $84\pm9$~km/s.   Comparing these values with the models with thin hydrogen layers of \citet{2020ApJ...901...93B} yields mass estimates of $1.11\pm0.04$~M$_\odot$ for  WD~0518-105 and $1.03\pm0.05$~M$_\odot$ for EGGR~25 in agreement with the results from the surface gravity and photometric radius measurements. The resolution of the spectrum of Lan~532 \citep{2011AJ....141...96L} is only 1.5~\AA\ which makes measuring the redshift of the spectrum somewhat difficult.  To obtain an estimate of the redshift, we cross-correlate the spectrum of the star with the best-fitting spectral model (as shown in Fig.~\ref{fig:Pleiad-Hlines}).  We obtain the value $77\pm28$~km/s, yielding a gravitational redshift of $78\pm28$~km/s, also in agreement with the model expectation of $81\pm6$~km/s.  The expected error for a given signal-to-noise is proportional to the square root of the resolution, so higher resolution spectra of Lan~532, EGGR 25 and WD~0518-105 would yield tighter constraints on the measured redshift and gravitational acceleration of these objects and test the white-dwarf atmosphere and interior models even further.

To estimate the mass of the progenitor stars from which these white dwarfs originated, we assume that the age of the Pleiades is 130~Myr \citep{2018A&A...616A...1G} with an uncertainty of 10~Myr and use the pre-white-dwarf lifetimes from \citet{2016ApJ...823..102C}.  The three Pleiad white dwarfs fill out the high-mass end of the diagram and increase the evidence for a break in the initial-final mass relation at a white-dwarf mass of about one solar mass and a progenitor mass of about four solar masses.  
\begin{figure}
    \centering
    \includegraphics[width=\columnwidth]{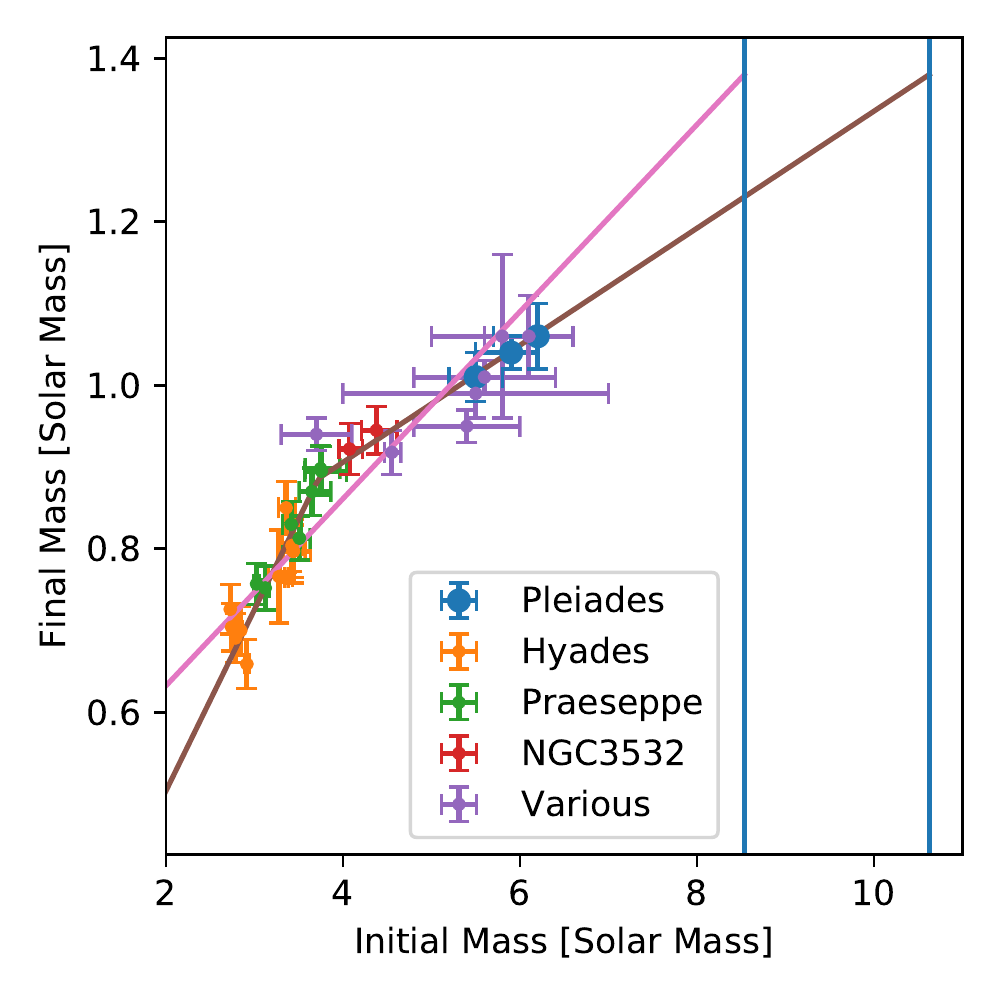}
    \caption{Initial-Final Mass Relation for White Dwarfs.  The vertical blue lines show the initial mass at which the white-dwarf mass would equal 1.38~M$_\odot$, the Chandrasekhar mass, for the two different fits. The Pleiades stars are from this paper. Those labeled from ``Various'' clusters are from \citet{2021ApJ...912..165R}, and the remainder are from \citet{2018ApJ...866...21C}.}
    \label{fig:Pleiad-IFMR}
\end{figure}

We fit the initial-final mass relation from white dwarfs within initial masses greater than two solar masses with a single linear fit,
\begin{eqnarray}
M_i =
 8.74 (M_f-0.5~\textrm{M}_\odot) + 1.72 \textrm{M}_\odot \\
M_f = 
 0.11 (M_i-1.72~\textrm{M}_\odot) + 0.5 \textrm{M}_\odot 
\end{eqnarray}
with $M_i(1.38\textrm{M}_\odot) = 8.5 \textrm{M}_\odot$.  We also perform 
a piece-wise linear fit
\begin{eqnarray}
M_i = \left \{ \begin{array}{lc}
 4.51 (M_f-0.5~\textrm{M}_\odot) + 1.98 \textrm{M}_\odot & M_f <  0.89 \textrm{M}_\odot \\
 13.97 (M_f-0.89~\textrm{M}_\odot) + 3.72 \textrm{M}_\odot & M_f \geq  0.89 \textrm{M}_\odot 
 \end{array} \right . \\
M_f = \left \{ \begin{array}{lc}
 0.22 (M_i-1.98~\textrm{M}_\odot) + 0.5 \textrm{M}_\odot & M_i <  3.72 \textrm{M}_\odot \\
 0.0716 (M_i-3.72~\textrm{M}_\odot) + 0.89 \textrm{M}_\odot & M_i \geq  3.72 \textrm{M}_\odot 
 \end{array} \right .
\end{eqnarray}
with a larger value for the mass of the progenitor corresponding to the Chandrasekhar mass, $M_i(1.38\textrm{M}_\odot) = 10.6 \textrm{M}_\odot$.
Given the uncertainties in the mass determinations, either relation provides an adequate fit ($\chi^2=4.33$ with 28 degrees of freedom or $\chi^2=2.00$ with 26 degrees of freedom), even though the values as plotted in Fig.~\ref{fig:Pleiad-IFMR} give a strong indication of a break in the IFMR.

\section{Conclusions}
\label{sec:conclusions}

The Pleiades star cluster has provided a benchmark for the characteristics and evolution of stars since prehistoric times.  Here we take a first look at the prehistory of the Pleiades by examining the properties and locations of its former members.  We have generated a catalogue of the escapee candidates from the Pleiades from the Gaia EDR3 database and focused on the positions, masses, colours and luminosities of these stars with a particular focus on the three white dwarfs that we associate with the Pleiades. Much work remains with the Pleiad escapee sample itself from verifying the technique on a star by star basis to understanding the dynamics of the evaporation of the Pleiades.  Many of the stars in the sample may have measured radial velocities besides those found in Gaia DR2 which could provide a confirmation of the reconstructed radial velocity for these stars and an additional check on the technique.  Furthermore, the colour-magnitude diagram of these escapees can be studied from the ultraviolet to the infrared with existing large surveys to probe this sample in further detail.  From a theoretical point of view, the sample can address the question of how cluster evaporation proceeds and whether external or internal factors play a larger role.

An orthogonal direction to the ideas above is to apply these techniques to other nearby clusters. \citet{2021ApJ...912..165R} have performed a search similar in spirit to the current one by looking for massive white dwarfs that escaped from more distant clusters.  In their search, the relative proper motion of the cluster and the stars is considered and constrained to be small, but the time for the star to reach its current position or the inferred radial velocity did not play a role in the analysis.  At these larger distances, uncertainties in proper motions and parallaxes are more important.  With these provisos in mind, future work could naturally be restricted to clusters within about 200~pc of the Sun to mitigate the effects of measurement uncertainties.  The distance restriction also yields a restriction on the age of the clusters themselves.  In particular a star travelling with a velocity of 1~km/s takes about 200~Myr to travel 200~pc; we would expect the earliest escapees from a 200~Myr-old cluster to lie more than 200~pc from the cluster today.  These two constraints yield a short list of clusters: the Pleiades, alpha Persei, NGC~2451A, IC~2391 and IC~2602.  These are the only clusters closer than 200~pc and also younger than 200~Myr.  We examine these clusters and re-examine the Pleiades in a companion paper \citep{youngclusters}.

\section*{Acknowledgements}

 This work was supported in part by NSERC Canada and Compute Canada.

This research has made use of the SIMBAD and Vizier databases, operated at CDS, Strasbourg, France and the Montreal White Dwarf Database produced and maintained by Prof. Patrick Dufour (Universit\'e de Montr\`eal) and Dr. Simon Blouin (LANL), 


This work has made use of data from the European Space Agency (ESA) mission
{\it Gaia} (\url{https://www.cosmos.esa.int/gaia}), processed by the {\it Gaia}
Data Processing and Analysis Consortium (DPAC,
\url{https://www.cosmos.esa.int/web/gaia/dpac/consortium}). Funding for the DPAC
has been provided by national institutions, in particular the institutions
participating in the {\it Gaia} Multilateral Agreement.

\section*{Data Availability}

The data used in this study was obtained from the ESA Gaia Archive using the commands outlined in the appendix and processed using TOPCAT.  The resulting catalogues and derived quantities are included in the catalogues.



\bibliographystyle{mnras}
\bibliography{main,parsec,yulwd}  




\appendix

\onecolumn

\section{Creating the Gaia EDR3 Pleiades Sample}

The sample was generated using the following ADQL query to the ESA Gaia archive:
\begin{verbatim}
SELECT * FROM gaiaedr3.gaia_source WHERE 
  (parallax>5 AND parallax_over_error>5 AND 1=CONTAINS(POINT('ICRS', ra, dec),
      CIRCLE('ICRS', 56.75, 24.1167, 45))) OR (parallax > 10 AND parallax_over_error > 5) 
\end{verbatim}
After obtaining the sample from the Gaia archive, the following commands in TOPCAT \citep{2005ASPC..347...29T} create the additional vectorial quantities used in the paper, the velocity ${\bf v}_{3D}$:
\begin{verbatim}
uvw3d icrsToGal(astromUVW(array(ra, dec, parallax, pmra, pmdec, dr2_radial_velocity)))
\end{verbatim}
the velocity of the star in the plane of the sky, ${\bf v}_{2D}$,
\begin{verbatim}
uvw2d icrsToGal(astromUVW(array(ra, dec, parallax, pmra, pmdec, 0)))
\end{verbatim}
and the position, ${\bf r}$
\begin{verbatim}
xyz icrsToGal(astromXYZ(ra, dec, parallax))
\end{verbatim}
All of these quantities have been converted to Galactic coordinates.
\section{Catalogue of current Pleiades stars}

The full catalogue is available by downloading the source for this submission from the arXiv.
{

\begin{longtable}{lrrrrrrrr}
\hline
       Gaia EDR3 Source ID                   &  \multicolumn{1}{c}{RA}     &  \multicolumn{1}{c}{Dec}     & \multicolumn{1}{c}{$G$}  &  \multicolumn{1}{c}{Parallax} & \multicolumn{1}{c}{$B_p-R_p$}  & \multicolumn{1}{c}{$\mu_\alpha$} & \multicolumn{1}{c}{$\mu_\delta$}  \\
                                        &  \multicolumn{1}{c}{[deg]}  &  \multicolumn{1}{c}{[deg]}   &  \multicolumn{1}{c}{[mag]}   &   \multicolumn{1}{c}{[mas]}   & \multicolumn{1}{c}{[mag]}  & \multicolumn{1}{c}{[mas/yr]} & \multicolumn{1}{c}{[mas/yr]}         \\
                            
\hline

     65283232316451328 &  56.4568 &  24.3675 & $  3.8633\pm  0.0027$ &  $  7.67\pm  0.31$ &  $ -0.0138\pm  0.0000$ &  $ 19.52\pm  0.45$ &  $-45.53\pm  0.28$ \\
    66529975427235712 &  57.2968 &  24.1365 & $  5.2033\pm  0.0009$ &  $  7.24\pm  0.13$ &  $ -0.0331\pm  0.0000$ &  $ 19.50\pm  0.15$ &  $-47.65\pm  0.10$ \\
    64940906245415808 &  57.0868 &  23.4210 & $  5.4283\pm  0.0008$ &  $  7.69\pm  0.10$ &  $ -0.0671\pm  0.0000$ &  $ 19.70\pm  0.11$ &  $-47.11\pm  0.08$ \\
    65287458566524928 &  56.2010 &  24.2893 & $  5.4413\pm  0.0009$ &  $  7.39\pm  0.07$ &  $ -0.0212\pm  0.0000$ &  $ 19.85\pm  0.08$ &  $-44.97\pm  0.06$ \\
    69812945346809600 &  56.2907 &  24.8391 & $  5.6404\pm  0.0006$ &  $  7.22\pm  0.06$ &  $ -0.0805\pm  0.0000$ &  $ 20.22\pm  0.07$ &  $-46.12\pm  0.04$ \\
    66798496781121792 &  56.4771 &  24.5543 & $  5.7520\pm  0.0006$ &  $  7.31\pm  0.08$ &  $ -0.0230\pm  0.0000$ &  $ 20.33\pm  0.09$ &  $-46.02\pm  0.06$ \\
    64053566003097984 &  57.4796 &  22.2440 & $  6.0404\pm  0.0006$ &  $  7.41\pm  0.10$ &  $  0.0235\pm  0.0000$ &  $ 21.22\pm  0.22$ &  $-42.77\pm  0.10$ \\
    66453490649681280 &  57.4315 &  23.7116 & $  6.1592\pm  0.0003$ &  $  7.46\pm  0.05$ &  $ -0.0497\pm  0.0001$ &  $ 17.87\pm  0.06$ &  $-45.57\pm  0.03$ \\
    66715105696289024 &  56.8378 &  24.1161 & $  6.2988\pm  0.0004$ &  $  7.20\pm  0.06$ &  $  0.0094\pm  0.0000$ &  $ 20.34\pm  0.07$ &  $-45.08\pm  0.04$ \\
    66786505232432512 &  56.5122 &  24.5277 & $  6.4253\pm  0.0004$ &  $  7.42\pm  0.04$ &  $ -0.0169\pm  0.0001$ &  $ 19.58\pm  0.05$ &  $-44.88\pm  0.03$ \\
    66745617143891968 &  57.3407 &  24.3808 & $  6.6049\pm  0.0003$ &  $  7.19\pm  0.04$ &  $ -0.0326\pm  0.0001$ &  $ 18.57\pm  0.05$ &  $-45.35\pm  0.03$ \\
    64679840952155264 &  54.6698 &  22.6594 & $  6.7019\pm  0.0003$ &  $  7.37\pm  0.04$ &  $  0.0001\pm  0.0001$ &  $ 20.16\pm  0.05$ &  $-45.36\pm  0.03$ \\
    66789872486779520 &  56.7476 &  24.5199 & $  6.8028\pm  0.0003$ &  $  7.27\pm  0.04$ &  $  0.0346\pm  0.0001$ &  $ 22.27\pm  0.04$ &  $-47.52\pm  0.03$ \\
    66507469798631808 &  57.4920 &  23.8485 & $  6.8064\pm  0.0003$ &  $  7.39\pm  0.04$ &  $  0.0553\pm  0.0001$ &  $ 18.38\pm  0.05$ &  $-45.87\pm  0.03$ \\
    66729880383767168 &  56.8728 &  24.2881 & $  6.8159\pm  0.0006$ &  $  7.23\pm  0.05$ &  $  0.1014\pm  0.0000$ &  $ 18.55\pm  0.06$ &  $-46.92\pm  0.03$ \\
    64956127609464320 &  56.4535 &  23.1469 & $  6.8952\pm  0.0003$ &  $  7.42\pm  0.04$ &  $  0.0246\pm  0.0001$ &  $ 20.51\pm  0.05$ &  $-45.70\pm  0.03$ \\
    66725379258016768 &  57.1255 &  24.3453 & $  6.9398\pm  0.0003$ &  $  6.90\pm  0.22$ &  $  0.1819\pm  0.0001$ &  $ 19.61\pm  0.27$ &  $-45.36\pm  0.17$ \\
    65017627244152064 &  56.8375 &  23.8031 & $  6.9976\pm  0.0003$ &  $  7.31\pm  0.04$ &  $  0.0424\pm  0.0001$ &  $ 19.10\pm  0.04$ &  $-45.52\pm  0.03$ \\\end{longtable}

}
\begin{landscape}
\section{Catalogue of Pleiades-escapee candidates}
The full catalogue is available by downloading the source for this submission from the arXiv.

{\small

\begin{longtable}{rrrrrrrrrrrrrr}
\hline
        \multicolumn{1}{l}{Gaia EDR3 Source ID}                   &  \multicolumn{1}{c}{RA}     &  \multicolumn{1}{c}{Dec}     & \multicolumn{1}{c}{$G$}  &  \multicolumn{1}{c}{Parallax} & \multicolumn{1}{c}{$B_p-R_p$}  & \multicolumn{1}{c}{$\mu_\alpha$} & \multicolumn{1}{c}{$\mu_\delta$}  
       & \multicolumn{1}{c}{$\Delta v_{2D}$} &  \multicolumn{1}{c}{$d_\textrm{present}$} & \multicolumn{1}{c}{$v_r$} &  \multicolumn{1}{c}{$d_\textrm{min}$} & $\Delta v_\textrm{3D}$ &  $t_\textrm{escape}$ \\
                                        &  \multicolumn{1}{c}{[deg]}  &  \multicolumn{1}{c}{[deg]}   &  \multicolumn{1}{c}{[mag]}   &   \multicolumn{1}{c}{[mas]}   & \multicolumn{1}{c}{[mag]}  & \multicolumn{1}{c}{[mas/yr]} & \multicolumn{1}{c}{[mas/yr]}  
                                        &  \multicolumn{1}{c}{[km/s]}   &  \multicolumn{1}{c}{[pc]}                 & \multicolumn{1}{c}{[km/s]} & \multicolumn{1}{c}{[pc]} & \multicolumn{1}{c}{[km/s]} & \multicolumn{1}{c}{[Myr]}\\
                            
\hline

   163841083812606592 &  61.6518 &  27.5997 & $  5.1445\pm  0.0006$ &  $  8.07\pm  0.11$ &  $ -0.1494\pm  0.0000$ &  $ 21.15\pm  0.14$ &  $-51.79\pm  0.07$ &   1.23 &   17.4 &    3.4 & 10.9 & 2.16 &  6.1\\
   119006092005322112 &  51.2042 &  28.6523 & $  7.1103\pm  0.0004$ &  $  9.74\pm  0.04$ &  $  0.1496\pm  0.0001$ &  $ 29.26\pm  0.05$ &  $-57.95\pm  0.04$ &   0.85 &   35.8 &    0.4 &  9.2 & 2.29 & 14.8\\
    62890175323180928 &  50.0887 &  23.6890 & $  7.5496\pm  0.0003$ &  $  9.62\pm  0.03$ &  $  0.3076\pm  0.0001$ &  $ 29.63\pm  0.03$ &  $-58.04\pm  0.02$ &   0.14 &   34.2 &    4.5 &  1.3 & 0.33 & 100.3\\
    62413988007668352 &  49.4574 &  22.8320 & $  7.5794\pm  0.0002$ &  $  7.48\pm  0.02$ &  $  0.3884\pm  0.0003$ &  $ 23.07\pm  0.03$ &  $-44.76\pm  0.02$ &   0.29 &   16.0 &    5.0 &  6.9 & 0.29 & 48.0\\
    65292234570088064 &  56.1073 &  24.3945 & $  8.1218\pm  0.0003$ &  $  8.03\pm  0.02$ &  $  0.3834\pm  0.0005$ &  $ 21.95\pm  0.03$ &  $-49.20\pm  0.02$ &   0.19 &   11.2 &    3.4 &  1.2 & 2.35 &  4.7\\
   136347211441563136 &  43.5371 &  32.3360 & $  8.1518\pm  0.0004$ &  $  7.23\pm  0.04$ &  $  0.3007\pm  0.0004$ &  $ 24.58\pm  0.03$ &  $-42.01\pm  0.03$ &   0.59 &   34.0 &   -1.0 & 12.8 & 0.61 & 50.7\\
  3229690616319232640 &  72.3796 &  -0.6318 & $  8.4444\pm  0.0001$ &  $  6.60\pm  0.02$ &  $  1.1659\pm  0.0013$ &  $ 11.91\pm  0.02$ &  $-36.06\pm  0.02$ &   2.09 &   73.6 &   21.6 &  4.8 & 2.33 & 30.8\\
    70347033119553408 &  56.3126 &  26.8913 & $  8.5493\pm  0.0004$ &  $  6.55\pm  0.03$ &  $  0.3565\pm  0.0005$ &  $ 17.60\pm  0.03$ &  $-39.85\pm  0.02$ &   0.69 &   18.3 &    6.3 &  1.1 & 1.92 &  9.3\\
   479937379577969792 &  79.7768 &  64.5393 & $  8.8540\pm  0.0002$ &  $  6.08\pm  0.02$ &  $  0.4938\pm  0.0017$ &  $  7.47\pm  0.01$ &  $-36.47\pm  0.01$ &   1.35 &  113.5 &  -11.9 & 13.6 & 1.66 & 66.4\\
    71296667568555904 &  54.0353 &  27.3428 & $  9.2155\pm  0.0005$ &  $  7.78\pm  0.02$ &  $  0.7591\pm  0.0005$ &  $ 21.97\pm  0.03$ &  $-47.90\pm  0.02$ &   0.26 &   11.8 &    3.6 &  4.6 & 0.33 & 31.9\\
     3763976194792448 &  45.2768 &   5.0064 & $  9.2792\pm  0.0002$ &  $  7.95\pm  0.02$ &  $  0.7205\pm  0.0012$ &  $ 24.53\pm  0.02$ &  $-45.73\pm  0.02$ &   2.30 &   50.9 &   12.4 &  7.7 & 2.30 & 21.4\\
  2884080885842578176 &  89.4617 & -38.0676 & $  9.3530\pm  0.0011$ &  $ 13.75\pm  0.04$ &  $  0.9150\pm  0.0002$ &  $  5.81\pm  0.05$ &  $-19.37\pm  0.05$ &   1.13 &  129.4 &   32.2 &  9.8 & 1.15 & 109.8\\
    64317998547641344 &  55.6001 &  21.4733 & $  9.7122\pm  0.0002$ &  $  7.84\pm  0.02$ &  $  0.7113\pm  0.0023$ &  $ 21.14\pm  0.02$ &  $-47.94\pm  0.02$ &   0.39 &   10.5 &    6.3 &  5.4 & 0.87 & 10.1\\
   123561334319267328 &  48.6034 &  31.8804 & $  9.9358\pm  0.0012$ &  $  9.09\pm  0.02$ &  $  0.8191\pm  0.0003$ &  $ 27.65\pm  0.02$ &  $-54.51\pm  0.02$ &   0.81 &   34.2 &   -0.5 &  8.6 & 1.14 & 28.3\\
  2569833036623574912 &  30.9437 &   9.0854 & $  9.9546\pm  0.0003$ &  $  7.78\pm  0.05$ &  $  0.7538\pm  0.0039$ &  $ 29.39\pm  0.07$ &  $-43.64\pm  0.05$ &   1.60 &   66.1 &    6.4 &  7.6 & 1.62 & 39.7\\
    59661291925385600 &  48.9759 &  20.4540 & $  9.9612\pm  0.0006$ &  $  9.78\pm  0.02$ &  $  0.8419\pm  0.0009$ &  $ 29.51\pm  0.02$ &  $-59.11\pm  0.02$ &   0.89 &   37.3 &    4.6 &  1.9 & 1.75 & 20.8\\
\end{longtable}

}
\bsp	
\label{lastpage}
\end{landscape}

\end{document}